\documentclass[twocolumn,secnumarabic,amssymb, nobibnotes, aps, prb]{revtex4-1}
\usepackage{graphicx}
\usepackage{epsfig}
\usepackage{amsmath,amssymb,amsfonts}
\usepackage{psfrag}
\usepackage{enumitem}
\usepackage{color}
\usepackage[colorlinks]{hyperref}
\usepackage{epstopdf}
\usepackage{tikz}

\begin{document}

\title{Instability of flux flow and production of vortex-antivortex pairs by current-driven Josephson vortices in layered superconductors }

\author{Ahmad Sheikhzada}
\email{asheikhz@odu.edu}
\author{Alex Gurevich}
\email{gurevich@odu.edu}

\affiliation{Department of Physics and Center 
for Accelerator Science, Old Dominion University, Norfolk, VA 23529, USA}


\begin{abstract}

We report numerical simulations of the nonlinear dynamics of Josephson vortices driven by strong dc currents in layered superconductors. Dynamic equations for interlayer phase differences in a stack of coupled superconducting layers were solved to calculate a drag coefficient $\eta(J)$ of the vortex as a function of the perpendicular dc current density $J$. It is shown that Cherenkov radiation produced by a moving vortex causes significant radiation drag increasing $\eta(v)$ at high vortex velocities $v$ and striking instabilities of driven Josephson vortices moving faster than a terminal velocity $v_c$. The steady-state flux flow breaks down at $v>v_c$ as the vortex starts producing a cascade of expanding vortex-antivortex pairs evolving into either planar macrovortex structures or branching flux patterns propagating both along and across the layers. This vortex-antivortex pair production triggered by a rapidly moving vortex is most pronounced in a stack of underdamped planar junctions where it can occur at $J>J_s$ well below the interlayer Josephson critical current density.  Both $v_c$ and $J_s$ were calculated as functions of the quasiparticle damping parameter, and the dc magnetic field applied parallel to the layers. The effects of vortex interaction on the Cherenkov instability of moving vortex chains and lattices in annular stacks of Josephson junctions were  considered. It is shown that a vortex driven by a current density $J>J_s$ in a multilayer of finite length excites self-sustained large-amplitude standing waves of magnetic flux, resulting in temporal oscillations of the total magnetic moment. We evaluated a contribution of this effect to the power $W$ radiated by the sample and showed that $W$ increases strongly as the number of layers increases. These mechanisms can result in nonlinearity of the c-axis electromagnetic response and contribute to THz radiation from  the layered cuprates at high dc current densities flowing perpendicular to the ab planes.

\end{abstract}

\maketitle
\section{Introduction}
\label{sec:intro}
The physics of current-driven Josephson (J) vortices  \cite{BP,KL} and its manifestations in flux flow oscillators\cite{ffo1,ffo2,ffo3}, THz radiation sources\cite{thz1,thz2,thz3,thz4}, nanoscale superconducting structures for digital memory \cite{qc,jm} current transport through grain boundaries\cite{HM,D,physc} in superconducting polycrystals and radio-frequency superconducting cavities for particle accelerators \cite{ag_srf}, have been areas of active experimental and theoretical investigations. Particularly, dynamics of J vortices in layered superconductors has attracted much attention since the discoveries of the cuprate and iron-based superconductors which exhibit an intrinsic Josephson effect between weakly coupled $ab$ planes \citep{kliener92,csg1,klcsg,yugens}. Numerical simulations of stacks of  Josephson junctions (JJ) have revealed instabilities of sliding Josephson vortex lattices \cite{kl-ch,jvlins,kosh} which affect the power of coherent THz radiation from single crystal BSCCO mesas  \cite{kosh,machidarec,tachiki,radcal,recjvl,insrecjvl}. New imaging tools have probed vortices at nanometer scales and revealed hypersonic vortices moving much faster than the velocity of superfluid condensate \cite{embon}.   

It has been usually assumed that a driven vortex preserves its identity as a topological defect no matter how fast it moves, because instability of a vortex would violate the fundamental conservation of the winding number $n=\pm 1$ in the superconducting order parameter $\Psi=\Delta\exp(in\chi)$. One of the outstanding questions is whether this topologically protected stability of a moving vortex remains preserved at any current below the depairing limit or there is a terminal velocity above which a uniformly moving vortex cannot exist. As far as the Josephson vortices are concerned, numerical simulations of long underdamped junctions\cite{screp}, planar JJ arrays \cite{bob,nakajima,paco} and a few coupled JJs \cite{a1,a2,a3,a4,a5}, and discrete sine-Gordon systems\cite{sg1,sg2}, have shown that there is indeed a terminal velocity $v_c$ above which uniform motion of a vortex driven by a dc current breaks down due to Cherenkov radiation.  The Cherenkov radiation of a vortex moving with a constant velocity $v$ is characteristic of high-$J_c$ Josephson junctions (JJ) or arrays of coupled JJs  in which the phase velocity of electromagnetic waves $v_p(k)$ decreases as the wave number $k$ increases \cite{sakai,csg-td,kleiner2,ngai,miccsg,lin-sust}, so that the Cherenkov condition $v>v_p(k)$ can be more easily satisfied at short wavelengths.  The resulting Cherenkov wake behind a moving J vortex causes a significant radiation drag in addition to the conventional quasiparticle viscous drag \cite{a2}. It turns out that the steady-state motion of a J vortex in which the Lorentz force is balanced by the viscous and radiation drag forces can only be sustained at $v<v_c$. A vortex moving with a velocity $v>v_c$ starts producing a cascade of expanding vortex-antivortex (V-AV) pairs which form dynamic dissipative patterns \cite{screp,paco}. Such resistive transition can occur at current densities $J>J_s$ which can be well below the critical current density of the interlayer junction $J_0$. Generation of V-AV pairs by a moving vortex pertains to a broader issue of stability of driven topological defects that can destroy global long range order in a way similar to the crack propagation resulting from the pileup of dislocations of opposite polarity \cite{disl}. Such process was observed in simulations of vortices in long JJs and planar JJ arrays where driven vortices cause propagating phase cracks in superconducting long range order \cite{screp,paco}. 

A question whether a fast Josephson vortex can initiate the V-AV pair production in layered superconductors is of interest to the theory of nonlinear flux flow of vortices along the $ab$ planes in high-$T_c$ cuprates and pnictides or artificial multilayer structures. For instance, revealing the materials parameters which control the values of $v_c$ and $J_s$ are essential for understanding the high-field electromagnetic response along the c-axis.  Another issue pertains to dynamic dissipative structures which appear due to the V-AV chain reaction triggered by a single moving vortex. The nonlinear dynamics of these structures and their effect on the radiation and other electromagnetic properties of layered superconductors are of particular interest.  The Cherenkov instability of vortices at high velocities is facilitated in underdamped interlayer junctions, as characteristic of highly anisotropic Bi-based cupraes, which can thus be testbeds for the experimental and theoretical investigations of these issues.       

The effects of Cherenkov radiation on a current-driven vortex in a few coupled junctions \cite{a1,a2,a3,a4,a5} or structural instabilities of driven vortex lattices and their manifestations in the THz radiation sources  \cite{kosh,kl-ch,jvlins} have been thoroughly investigated. Yet little is known about dynamics of macrovortex flux structures resulting from the V-AV pair production caused by a driven J vortex in multilayered superconductors. In this work we address this issue, including a nonlinear vortex viscosity controlled by the ohmic and radiation drag, and the factors determining the terminal velocity $v_c$ and the threshold critical current density $J_s$ at which the steady state flux flow breaks down. We investigate spontaneous generation of V-AV pairs by a moving vortex at $v>v_c$ and show that they result in macrovortex structures spreading both along and across the layers. It turns out that in a stack of underdamped JJs of finite length the V-AV pair production caused by a vortex shuttle excites  large-amplitude standing waves of magnetic flux, giving rise to oscillations in the total magnetic moment and magneto-dipole radiation from the sample.  In our simulation we used the well-established equations that describe J vortices in layered superconductors modeled as a stack of planar JJs coupled by inductive currents and charging effects \cite{sakai,csg-td,kleiner2,ngai,miccsg,lin-sust}.
        
The paper is organized as follows. Sec. \ref{sec:elec} specifies the geometry of the problem and the equations used in numerical simulations. In Sec. \ref{sec:ch} we discuss  Josephson plasmons and conditions of Cherenkov radiation in layered superconductors. Sec. \ref{sec:sv} contains the results of our calculations of a nonlinear drag coefficient, terminal velocity and critical current density of the Cherenkov instability $J_s$ for a single vortex. It is shown that the production of V-AV pairs at $J>J_s$ results in branching dynamic patterns and macrovortex structures. In Sec. \ref{sec:vc} and \ref{sec:vl} we address the effects of vortex interaction on the Cherenkov instability of moving vortex chains and lattices in annular JJ stacks. In Sec \ref{sec:finsize} we consider dynamics of bouncing macrovortices and self-sustained flux standing waves of large amplitude excited by a V-AV shuttle in a JJ stack of finite length. Contribution of this effect to the power $W$ radiated by the JJ stack, and a strong increase of $W$ with the number of layers are addressed. The conclusions and broader implications of our results are presented in Sec. \ref{sec:disc}.

\section{Coupled sine-Gordon Equations}
\label{sec:elec}

\begin{figure}
\includegraphics[trim={1mm 0mm 0mm 0mm},clip, width=\columnwidth ]{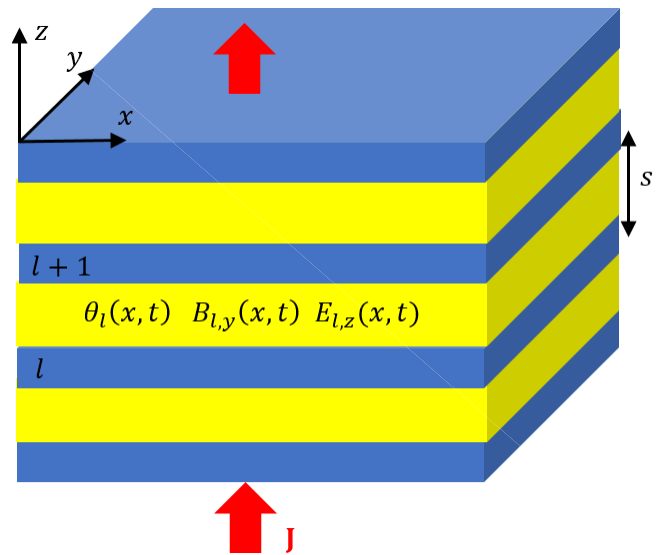}
\caption{Stack of intrinsic Josephson junctions (yellow) between superconducting layers (blue).}
\label{fig1}
\end{figure}

Consider vortices in a stack of long JJs between superconducting layers shown in Fig. \ref{fig1}. 
The dynamics of the phase difference $\theta_l(x,t)$ across the $l$-th junction, and the  
magnetic field $B_l(x,t)$ parallel to the layers can be described by the coupled sine-Gordon equations \cite{thz1,tachiki,sakai,csg-td,kleiner2,ngai,miccsg,lin-sust}
\begin{gather}
(1-\alpha\Delta_d)\theta_l''=\nonumber \\ (1-\zeta\Delta_d)[(1-\alpha\Delta_d)\sin\theta_l+\beta+\eta\dot{\theta_l}+\ddot{\theta_l}],
\label{csg} \\
B_l=(1-\zeta\Delta_d)^{-1}\theta_l'.
\label{B}
\end{gather}    
Here $\Delta_d f_l\equiv f_{l+1}+f_{l-1}-2f_l$ is the lattice Laplacian,
the prime and overdot denote partial derivatives with respect to the dimensionless coordinate $x/\lambda_c$ and time $\omega_J t$, respectively, $\omega_J=c/\sqrt{\epsilon_c} \lambda_c$ is the Josephson plasma frequency, $c$ is the speed of light, $\epsilon_c$ is the dielectric constant along the $z$ axis, $\lambda_c$ is the magnetic field penetration depth along the layers ($\textbf{B}$ parallel to the $ab$ planes in cuprates), and $B$ is measured in units of $\phi_0/2\pi s\lambda_c$ where $\phi_0$ is the flux quantum. The viscous drag coefficient $\eta$ and the dimensionless current $\beta$ are defined by: 
\begin{equation}
\eta=\frac{\sigma_c\lambda_c}{\epsilon_0\sqrt{\epsilon_c}c},\qquad \beta=\frac{J}{J_0},
\label{par1}
\end{equation}
where $J$ is the density of a uniform bias current flowing across the layers, $J_0$ is the critical current density of the junctions, $\sigma_c$ is the interlayer quasiparticle conductivity, and $\epsilon_0$ is the vacuum permittivity. The dimensionless damping parameter  $\eta$ in BSCCO crystals is typically $\simeq 0.005-0.05$\cite{lin-sust,machida}. The parameters $\alpha$ and $\zeta$ in Eq. (\ref{csg}) quantify charge and inductive coupling of the layers, respectively:
\begin{equation}
\alpha=\epsilon_cl_{TF}^2/s^2,\qquad\zeta=(\lambda_{ab}/s)^2.
\label{prmtr}
\end{equation}
Here $l_{TF}$ is the Thomas-Fermi screening length along the layers, $\lambda_{ab}$ is the magnetic field penetration depth for $\textbf{B}$ parallel to the $c$ axis, and $s$ is the spacing between the superconducting layers. For a BSCCO crystal with the anisotropy parameter $\Gamma\equiv\lambda_c/\lambda_{ab}\sim 500$, $\lambda_{ab}\sim 400$ nm, $\lambda_c\sim 200$ $\mu$m and $s=1.5$ nm, $\zeta \sim 10^5$ is much larger than the typical value of $\alpha\sim 1$. In this case the term $\alpha\Delta_d$ which describes deviations from charge neutrality in Eq. (\ref{csg}) can be neglected \cite{lin-sust}, so that Eq. (\ref{csg}) reduces to:
\begin{equation}
\theta_l''=(1-\zeta\Delta_d)(\sin\theta_l+\beta +\eta\dot{\theta_l}+\ddot{\theta_l}).
\label{sge}
\end{equation}    
In this work we performed numerical simulations Eq. (\ref{sge}) using the method of lines \cite{mdln,mdabm}. Charging effects were neglected, unless specified otherwise. 

\section{Cherenkov radiation and instability}
\label{sec:ch}
 
Josephson vortices described by Eq. (\ref{sge}) have two length scales along the $xy$ planes: the length of the Josephson core $\lambda_J\equiv\Gamma s$ and the magnetic penetration depth $\lambda_c$ determining the scale of circulating currents along the stack. Equation (\ref{csg}) also describes small amplitude waves $\delta\theta \propto e^{ik_x x+iqz-i\omega t}$ ~ \cite{sakai,lin-sust}. If the number of layers $N\to\infty$, linearization of Eq. (\ref{csg}) with respect to $\delta \theta$ around the uniform current state $\sin\theta_0=-\beta$ yields the following dispersion relation $\omega(k_x,q)$ for the Josephson plasma waves (in the original units): 
\begin{gather}
\omega(k_x,q)=\Omega(k_x,q) - \frac{i\eta\omega_J}{2},
\label{omk} \\
\!\!\Omega^2=\bigl[(1+\alpha_q)\sqrt{1-\beta^2}-\frac{\eta^2}{4}\bigr]\omega_J^2
+\biggl[\frac{1+\alpha_q}{1+\zeta_q}\biggr](k_x c_i)^2,
\label{Om} \\
\alpha_q=4\alpha\sin^2\frac{qs}{2},\qquad \zeta_q=4\zeta\sin^2\frac{qs}{2},
\label{alze}
\end{gather}
where $c_i=\lambda_c\omega_J=c/\sqrt{\epsilon_c}$ is the speed of light in the dielectric layers. At $\eta\to 0$ and $k_x=q=0$ 
Eqs. (\ref{omk})-(\ref{alze}) yield $\omega=\omega_J(1-\beta^2)^{1/4}$ but at $\lambda_c k_x \gg 1$ the frequency of the Josephson plasmon $\omega (k_x,q)=\tilde{c}(q)k_x$ depends linearly on the in-plane wave number $k_x$. Here 
the longitudinal phase velocity $\omega/k_x=\tilde{c}(q)$ depends on the $z$-component $q$ of the wave vector:
\begin{equation}
\tilde{c}(q)=c_i\!\left[\frac{1+4\alpha\sin^2(qs/2)}{1+4\zeta\sin^2(qs/2)}\right]^{1/2}.
\label{swih}
\end{equation} 
For a stack of $N$ junctions, Eqs. (\ref{omk})-(\ref{alze}) with $q_n=\pi n/(N+1)s$ and $n=0,1, ... N$, describe 
$N+1$ branches of plasma waves\cite{lin-sust}.
In the case of $\zeta\gg\alpha$ characteristic of the layered cuprates, $\tilde{c}$ decreases strongly as $q$ increases, from $\tilde{c}=c_i$ at $q=0$ to 
$\tilde{c}= c_i/2\sqrt{\zeta}\ll c_i$ at $q=\pi/s$. Thus, the plasma wave with alternating $\theta_l$ in the $z$ direction has the minimum phase velocity $c_s=c_i/2\sqrt{\zeta}=cs/2\lambda_{ab}\sqrt{\epsilon_c}$ corresponding to the Swihart velocity in a single junction \cite{BP}. These features of $\Omega(k_x,q)$ give rise to Cherenkov radiation produced by a moving vortex  \cite{thz1,Kl,savelev,krasnov}. 

\begin{figure}
\includegraphics[width=\columnwidth]{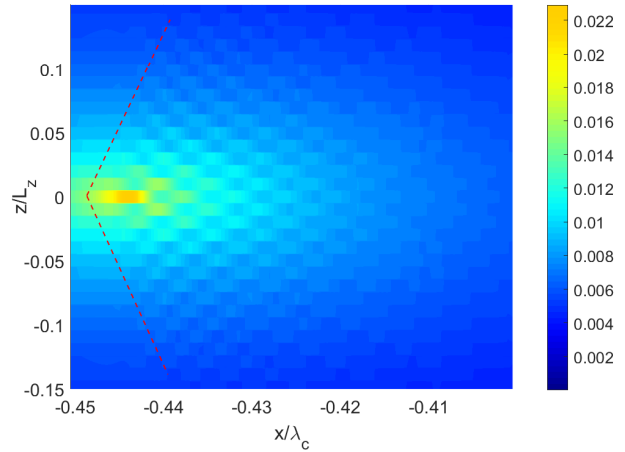}
\caption{Colormap of Cherenkov radiation cone in the magnetic field $B_l(x)$ produced by a vortex moving uniformly in the middle layer in a stack of  $N=101$ junctions. Here $B_l(x)$ is obtained by simulations of Eqs. (\ref{csg}) with $\beta=0.25$, $\zeta = 71111$, $\alpha= 1$, $\eta = 0.05$ and $B_0=\phi_0/2\pi s\lambda_c$. Only solutions for 15 neighboring junctions above and below the vortex are shown. Note that $L_z=Ns\sim 10^{-3}\lambda_c$ so the vortex is strongly elongated along the $x$ direction. }
\label{fig2}
\end{figure}

Cherenkov radiation occurs if the velocity $v$ of a vortex exceeds the minimum phase velocity $\Omega(k_x)/k_x$ of the Josephson plasmons. As follows from Eq. (\ref{swih}), the 
condition $v>\tilde{c}(q)$ at $(k_x\lambda_c)^2\gg 1$ and $\zeta\gg 1$ is first satisfied if $v>c_s$ at $q=\pi/s$.  For instance, Fig. \ref{fig2} shows the Cherenkov radiation cone behind a moving vortex obtained by numerical simulations of Eq. (\ref{csg}). 

\section{Single vortex}
\label{sec:sv}
\subsection{Laterally infinite stack} 

\begin{figure}
\includegraphics[width=\columnwidth]{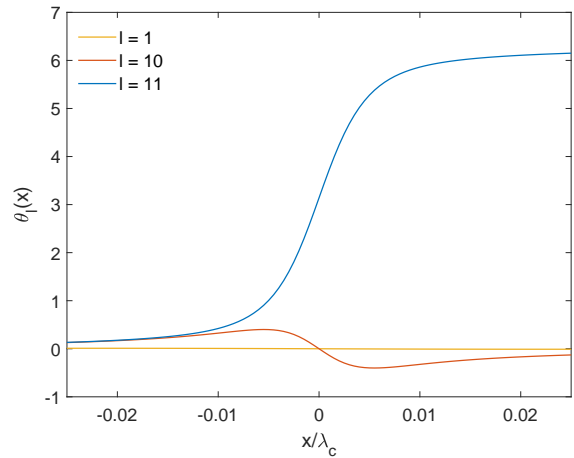}
\caption{Phase profile of a static vortex in the middle junction ($l=11$) and $\theta_l(x)$ induced by the vortex on the layers with $l=10$ and $l =1$). Here $\theta_l(x)$ are symmetric with respect to the central layer.}
\label{fig3}
\end{figure}

\begin{figure}
\includegraphics[width=\columnwidth]{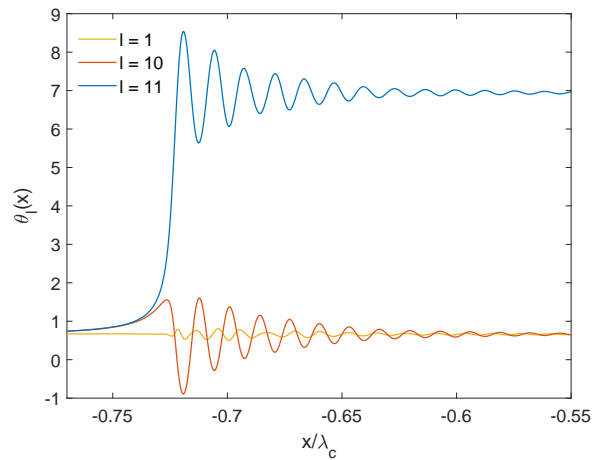}
\caption{Phase profiles of a single vortex propagating along the middle junction $(l=11)$ and the trailing tail of Cherenkov radiation produced on the neighboring junctions ($l = 1$ and $l=10$) calculated from Eq. (\ref{sge}) at $\beta=0.615$ and $\eta= 0.05$. }
\label{fig4}
\end{figure}

\begin{figure}
\includegraphics[width=\columnwidth]{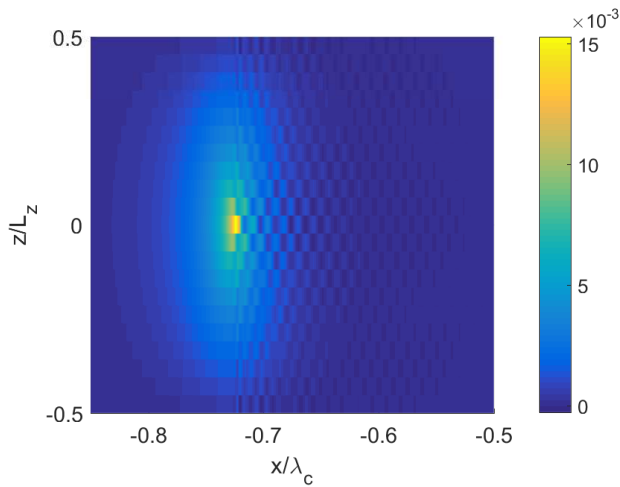}
\caption{A color map of the magnetic field in the vortex moving along the central junction calculated from Eq. (\ref{B}) at $\beta=0.615$ and $\eta = 0.05$. Here Cherenkov radiation behind the vortex manifests itself as color ripples. Since $L_z\sim 10^{-4}\lambda_c$, the vortex is strongly elongated along the $x$ direction.}
\label{fig5}
\end{figure}

In this section we present results of simulations of Eq. (\ref{sge}) describing vortices in a stack of $N=21$ junctions with $\eta=0.05$.  Solution of Eq. (\ref{sge}) for a  stationary vortex in the middle layer is shown in Fig. \ref{fig3}. As the bias current $\beta$ increases the vortex velocity $v(\beta)$ controlled by the drag of quasiparticle currents and radiational forces increases. Here the viscous drag dominates at small  $\beta$ for which the driving Lorentz force is balanced by the ohmic friction due to dissipative quasiparticle currents in the moving vortex \cite{clem}.  At $\beta\simeq 0.075$ the velocity exceeds the threshold, $v>c_s$ at which the vortex starts radiating Cherenkov waves. As $\beta$ further increases the amplitude and the wavelength of this Cherenkov wake increase and radiation spreads across the neighboring junctions. Figures \ref{fig4} and \ref{fig5} show the calculated phase and field profiles around the moving vortex at $\beta=0.615$. 

\begin{figure}
\includegraphics[width=\columnwidth]{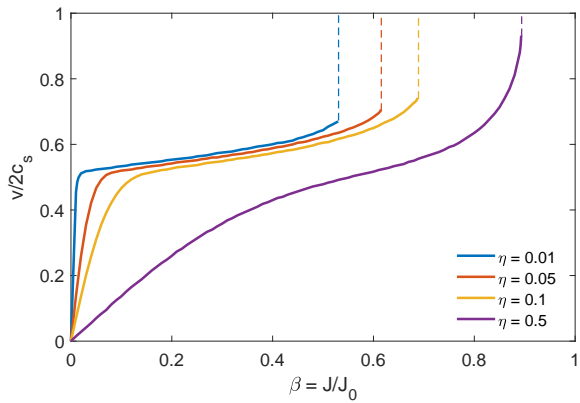}
\caption{Stationary velocities of a vortex moving along the central JJ as a function of the bias current at different $\eta$. The instability occurs at the endpoints of the curves. The sharp change in the slope of $v(\beta)$ at $\eta\ll 1$ indicates the transition from the ohmic to radiation vortex drag.}
\label{fig6}
\end{figure}

\begin{figure}
\includegraphics[width=\columnwidth]{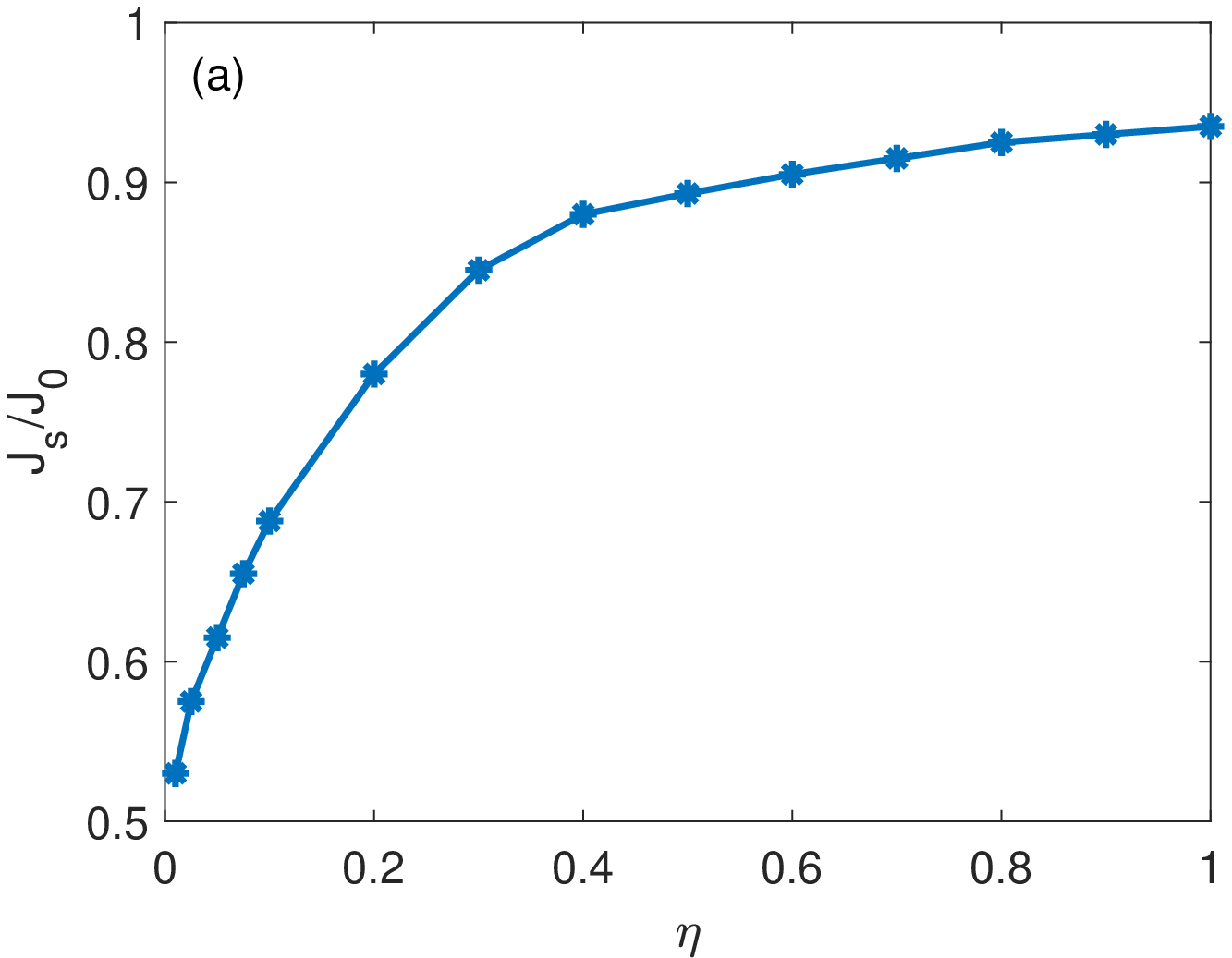}
\includegraphics[width=\columnwidth]{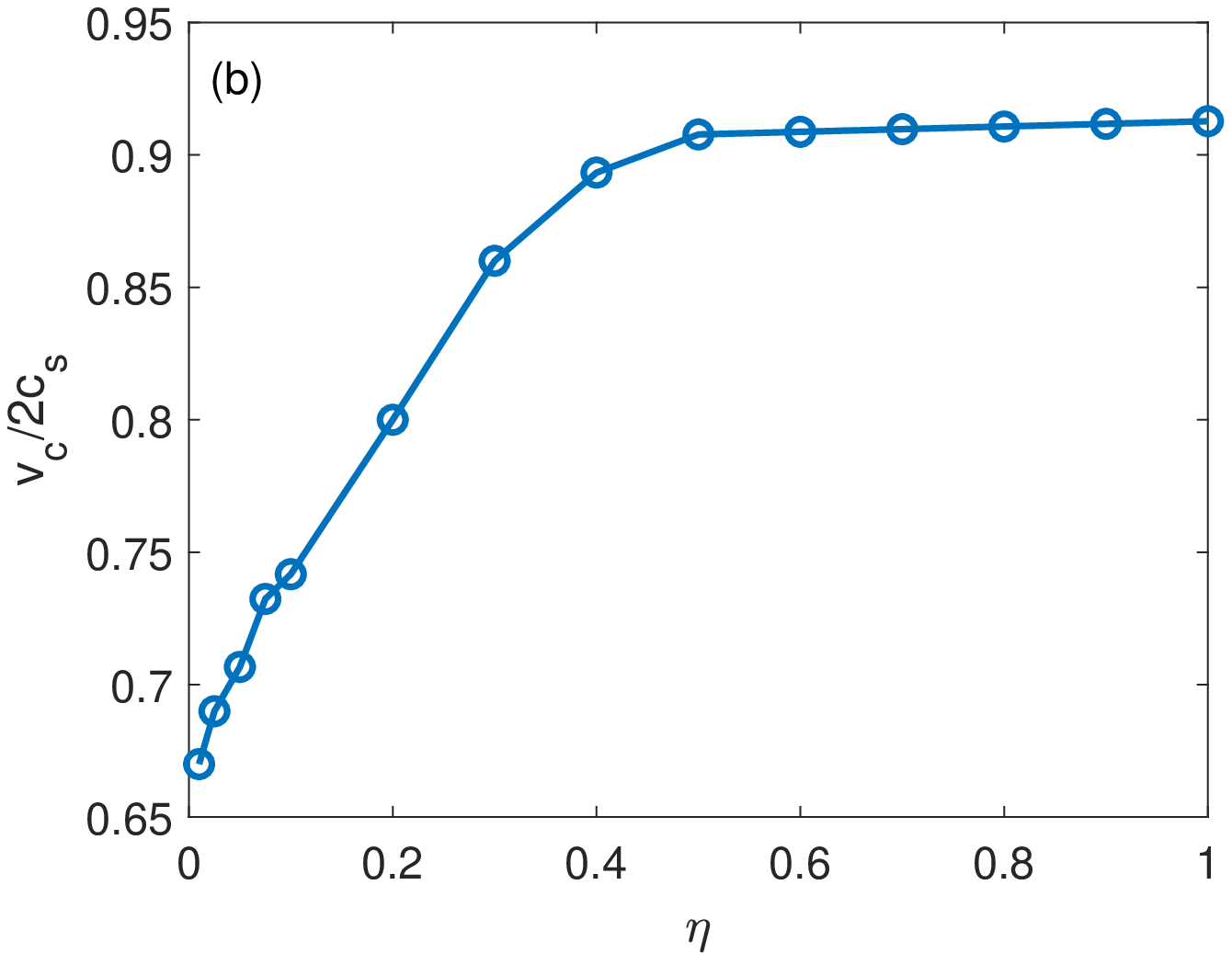}
\caption{The threshold instability current (a) and the terminal velocity (b) as functions of $\eta$ calculated for $\zeta=71111$. }
\label{fig7}
\end{figure}

Using the solutions for $\theta_l(x,t)$, we calculated the steady-state velocity of the vortex $v(\beta)$ as a function of the driving current $\beta$ at different values of $\eta$. The so-obtained curves $v(\beta)$ shown in Fig. \ref{fig6} have two distinct parts corresponding to different mechanisms of vortex drag.  At small currents the vortex velocity is limited by the quasiparticle viscous drag $dv/d\beta\propto \eta^{-1}$ and $v(\beta)$ increases sharply with $\beta$ if $\eta\ll 1$. The kink in the $v(\beta)$ curve at intermediate $\beta$ occurs at the onset of Cherenkov radiation above which the slope of $v(\beta)$ decreases as the radiation friction takes over \cite{thz1,mints} and $v(\beta)$ becomes weakly dependent on the dissipative term in Eq. (\ref{sge}). At $\eta\ll 1$ the radiation friction dominates at practically all $\beta$, significantly reducing $v(\beta)$ which exceeds the Cherenkov threshold. As $\eta$ increases the kink separating the ohmic and Cherenkov vortex drag regions of $v(\beta)$ gets less pronounced. All $v(\beta)$ curves have the endpoints at $\beta=\beta_s$ and $v=v_c$ beyond which Eq. (\ref{sge}) no longer has solutions for uniformly moving vortices. Figure \ref{fig7} shows the calculated critical current $\beta_s$ and the corresponding terminal vortex velocity $v_c$ as functions of the damping parameter $\eta$. For underdamped junctions $J_s(\eta)$ is well below $J_0$ and increases monotonically with $\eta$, approaching $J_0$ at $\eta > 1$. In turn, the terminal velocity increases from $v_c\approx 1.35c_s$ at $\eta\ll 1$ to $v_c \approx 1.85c_s$ at $\eta=1$. A similar behavior of $v(\beta)$ and $v_c$ was obtained previously by Goldobin et al. ~\cite{a2} in numerical simulations of two and three inductively coupled planar JJs.  

\begin{figure}
\includegraphics[width=\columnwidth]{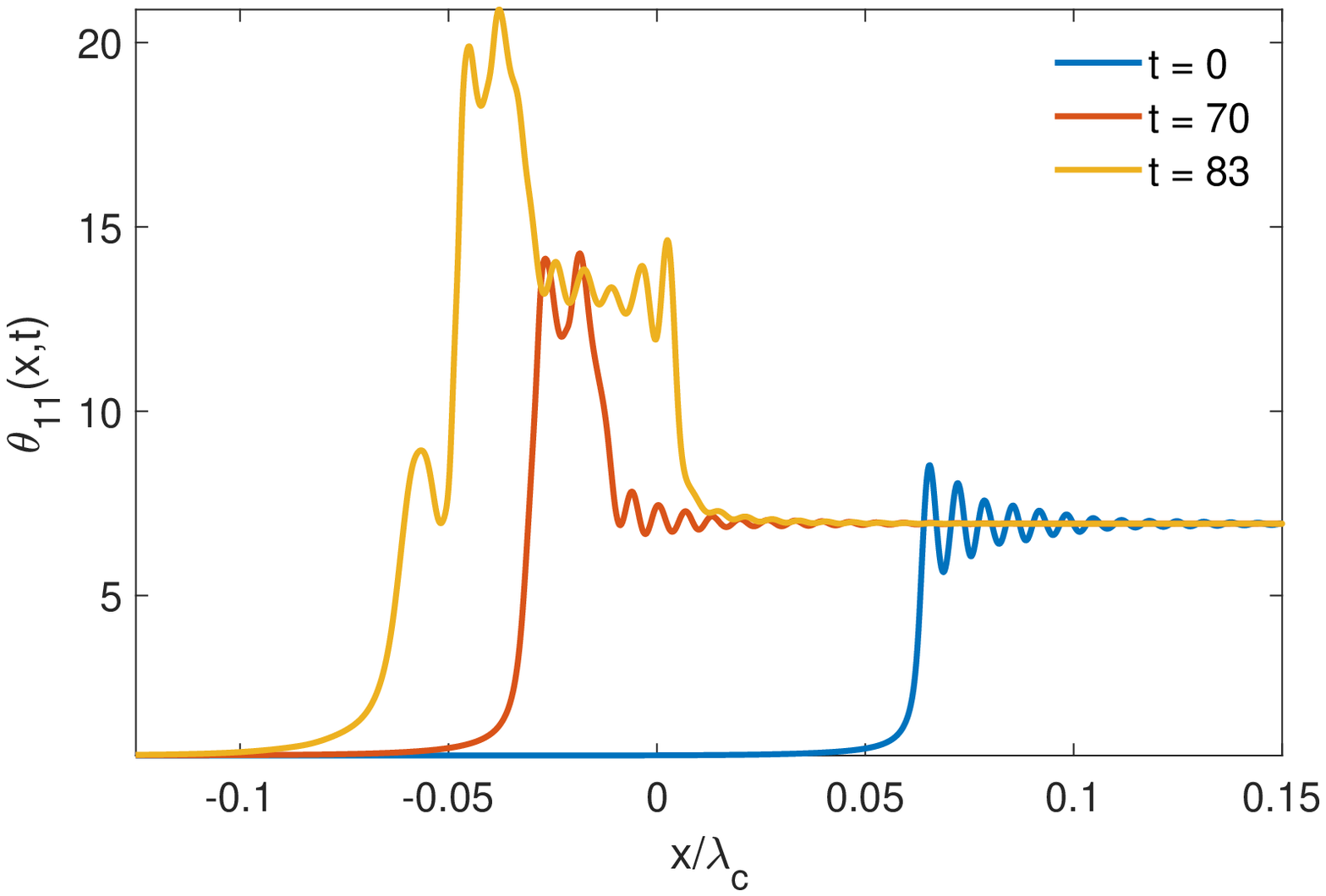}\\
\includegraphics[width=\columnwidth]{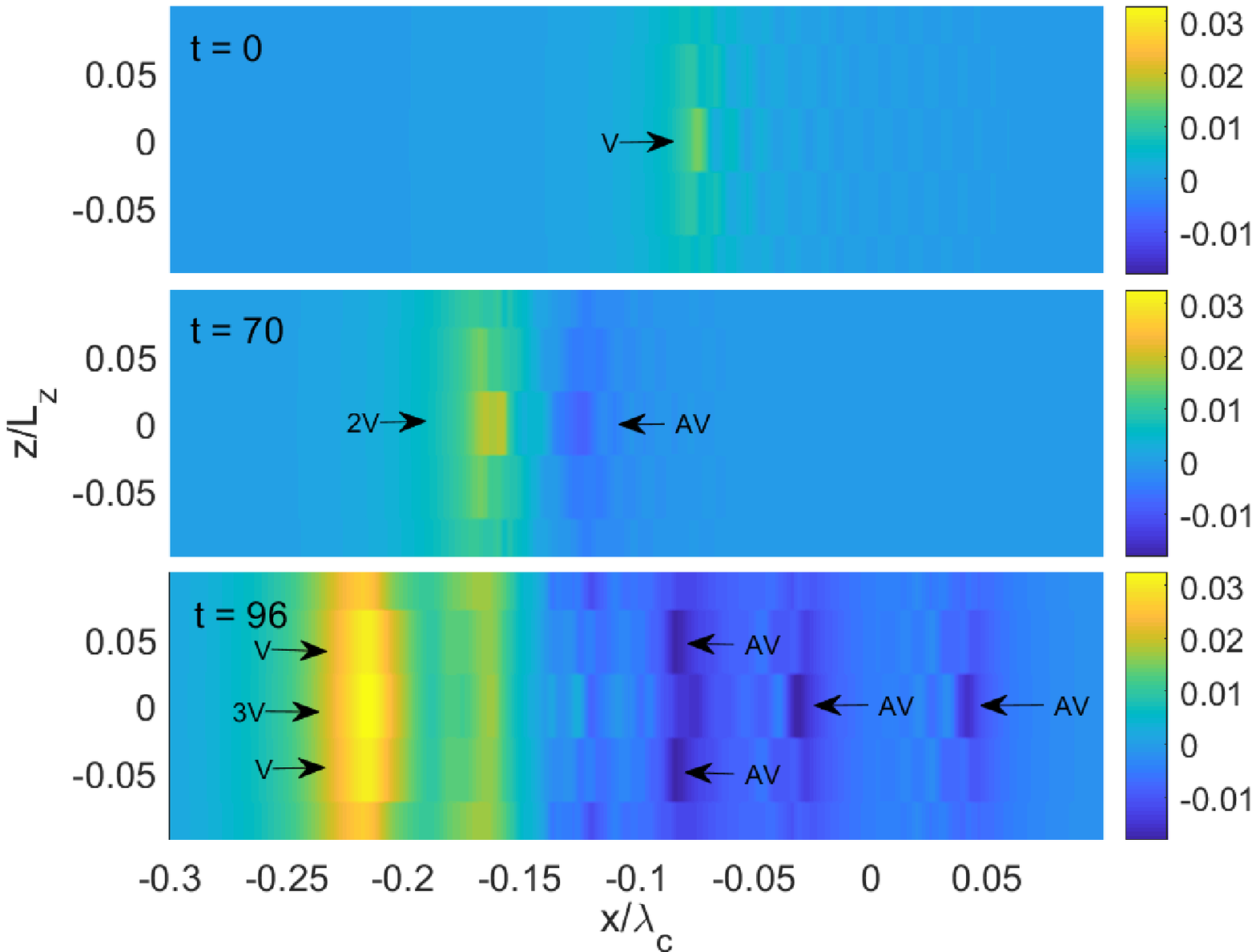}
\caption{Initial stages of generation of V-AV pairs by a vortex moving along the central junction (top panel), and snapshots of field distribution solutions showing the two dimensional growth of instability for junctions with $l=9,10$ and $11$ at three different times (bottom panel). The results are calculated at $\eta=0.05$, $\zeta=71111$ and $\beta=0.62$.}
\label{fig8}
\end{figure}

\begin{figure}
\includegraphics[width=\columnwidth]{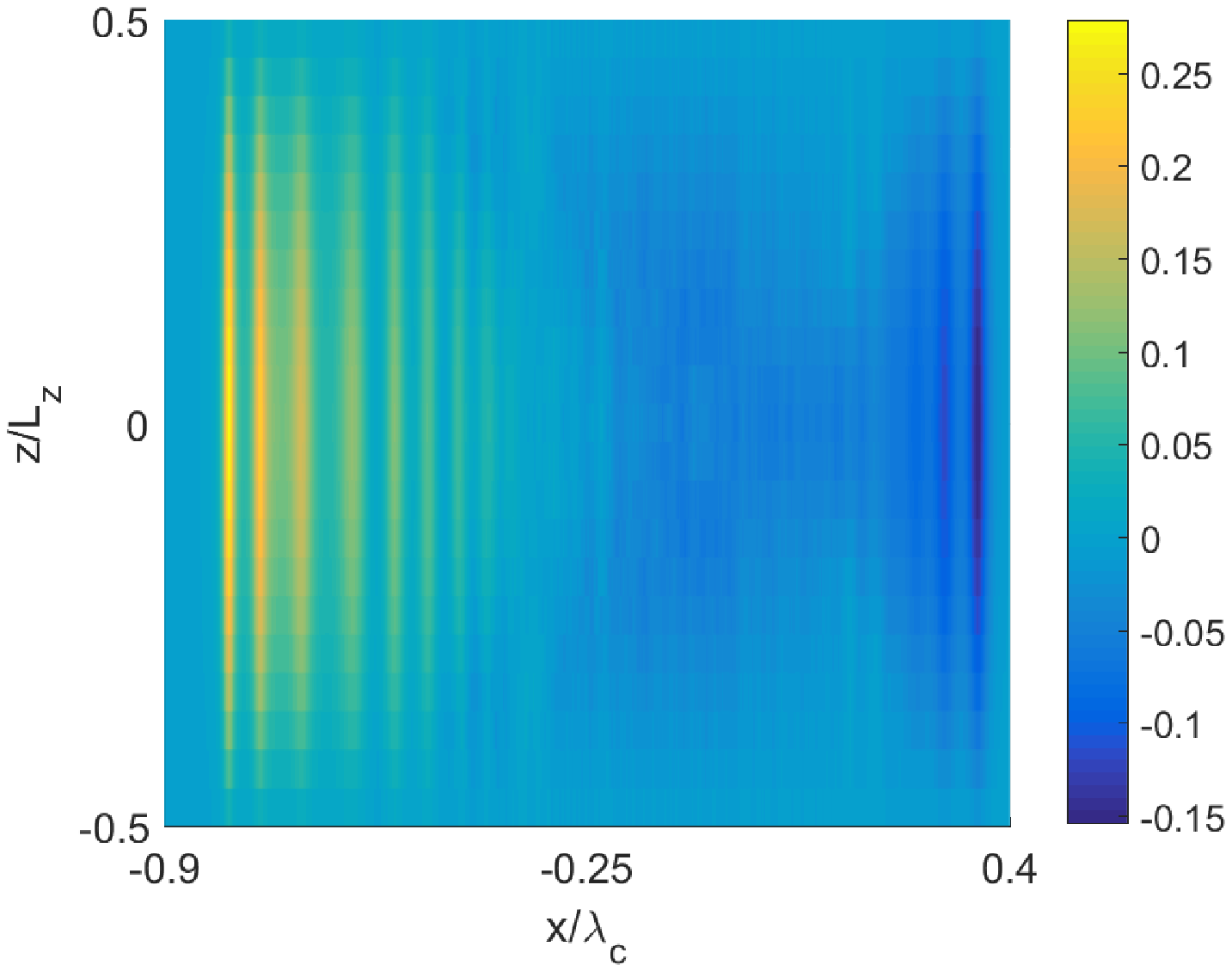}\\
\includegraphics[width=\columnwidth]{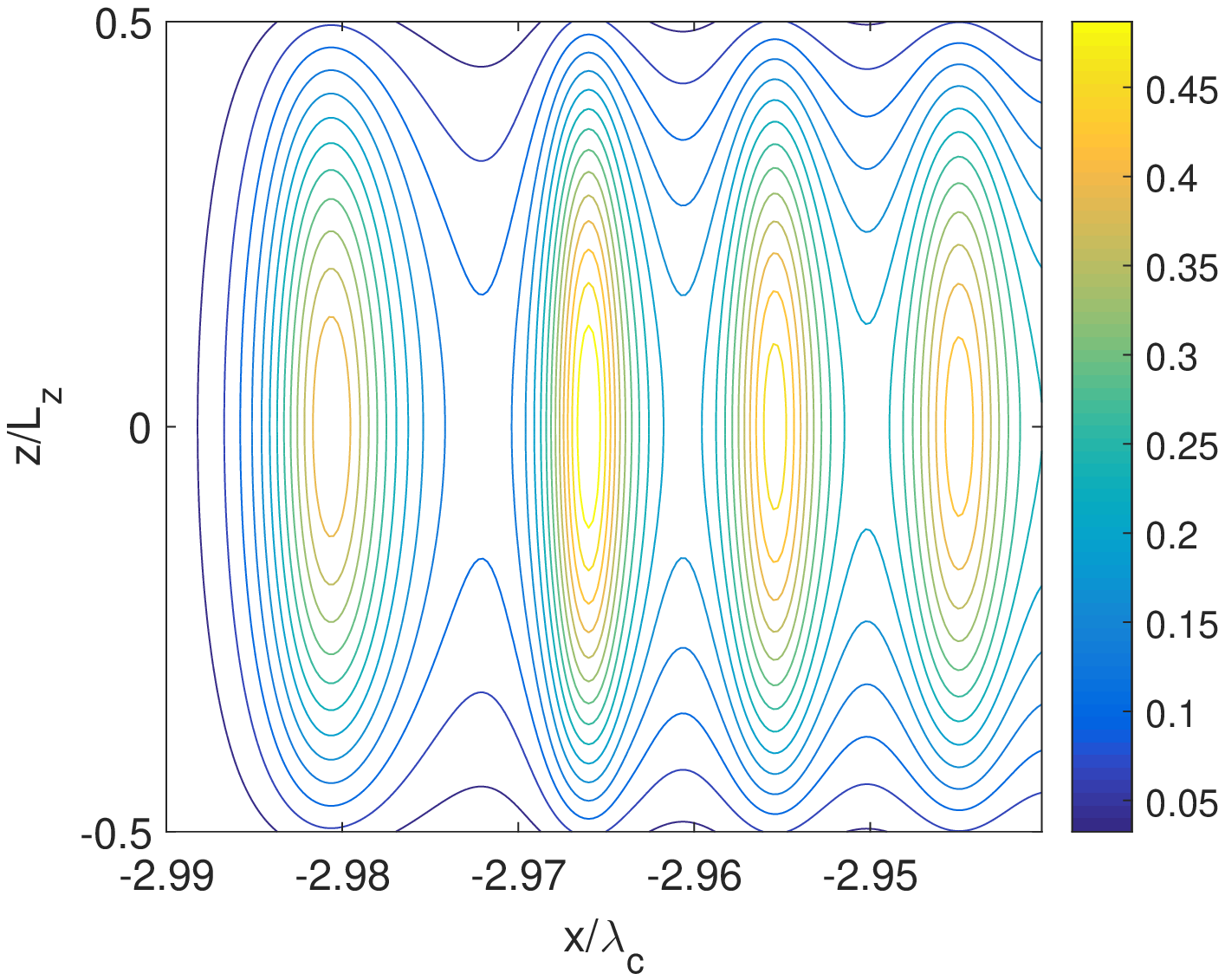}
\caption{Cross sectional view of the field distribution profiles in the stack after the instability (top panel, $t=125$) along with a close-up view of giant vortices moving to the left (bottom panel, $t=225$). Similar macro vortices with opposite polarity form at the other side of the stack (as shown in the top panel).}
\label{fig9}
\end{figure}

At $\beta>\beta_s$ in Eq. (\ref{sge}), the moving vortex starts spontaneously generating V-AV pairs which spread both along and across the JJ stack. For instance, at $\eta=0.05$ this process starts at $\beta_s\simeq 0.62$ and $v_c\approx \sqrt{2}c_s$. Such vortex splitting instability in a layered superconductor turned out to be similar to that of a driven vortex in a single JJ described by equations of nonlocal Josephson electrodynamics \cite{screp}.  This mechanism is illustrated by Fig. \ref{fig8} which shows that a critical nucleus being in the unstable $\pi-$phase state with $5\pi/2 <\theta<7\pi/2$ forms behind the vortex moving along the central layer where the maximum of Cherenkov radiation wake $\theta_l(x,t)$ reaches the threshold value $\theta_c\approx 8.6$.  As $\beta$ increases the amplitude and the width of this $\pi-$phase domain grows and eventually it splits, triggering a cascade of V-AV pairs which expand along the middle junction. In turn, the V-AV pairs in the middle junction induce V-AV pairs on the neighboring junctions which then start splitting and propagating along the layers and across the stack. This process produces an expanding chain of macrovortices which spread across the entire stack, the macrovortices of positive polarity accumulating at one edge of the stack while macrovortices of negative polarity accumulating at the other edge, as shown in Figs.  \ref{fig9}. A simulation video of this process is available in Ref. \onlinecite{supp}. 

The dynamics of the V-AV pair production caused by a single moving vortex, and the subsequent formation of the expanding macrovortex structure does not change qualitatively as the number of layers increases above $N=21$ used in the simulations described above.  For instance, our simulations for a stack with $N=101$ have shown that the V-AV pair production starts at $\beta=0.625$ which is very close to the instability current of a vortex in a stack with 21 junctions. Thus, the results obtained for $N=21$ can be representative of the BSCCO crystal mesas with $N\sim 1000$, consistent with the conclusion of Ref. \onlinecite{krasnov} that the behavior of vortices would become independent of the thickness of the stack if $N > \lambda_{ab}/s\sim 200$.

\subsection{Annular stack}

\begin{figure}
\includegraphics[width=\columnwidth]{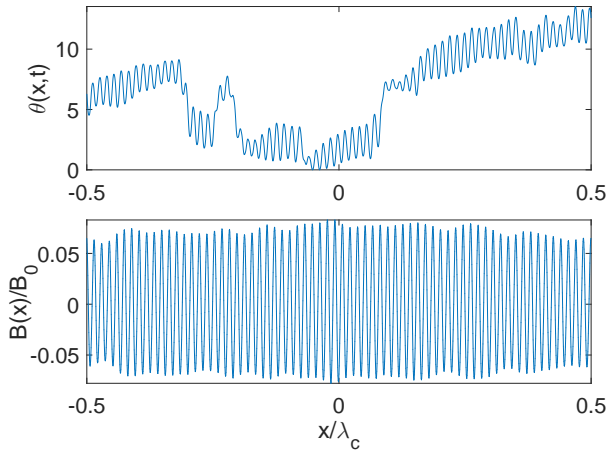}
\caption{Snapshots of representative solutions for $\theta_l(x,t)$ (top) and $B_l(x,t)$ (bottom) along the middle JJ at the critical current $\beta=\beta_s=0.62$.}
\label{fig10}
\end{figure}

\begin{figure}
\includegraphics[width=\columnwidth]{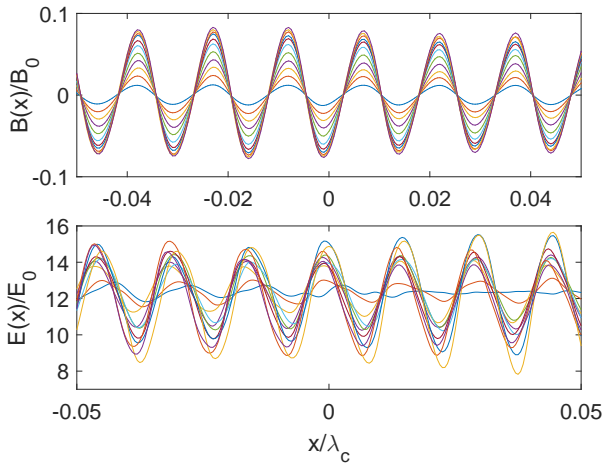}
\caption{Snapshots of the magnetic field (top) and electric field (bottom) in junctions 1-11 calculated at $\beta=\beta_s=0.62$, where $E_0=\phi_0\omega_J/2\pi c s$. Here the largest oscillation amplitude corresponds to the middle junction and the lowest amplitude corresponds to the top/bottom junction.}
\label{fig11}
\end{figure}

To investigate how the vortex dynamics changes by imposing the periodic boundary conditions, we consider an annular stack in which 
\begin{gather}
\theta_l(x=-L/2)=\theta_l(x=L/2)+ 2n\pi,\nonumber \\
\theta'_l(x=-L/2)=\theta'_l(x=L/2),
\label{bc}
\end{gather}
where $n=n_f-n_a$ is the difference of the number of fluxons $(n_f)$ and antifluxons $(n_a)$ on the $l-$th layer, and $L$ is the circumference of the stack along the $x$ direction. In our simulations we choose $L=\lambda_c\gg\lambda_J$ in which case the structure of a static vortex in the annular stack at $\beta=0$ is nearly identical to the vortex in the infinite stack shown in Fig. \ref{fig3}. If a transport current flows across the annular stack, a vortex moving along the central junction radiates Cherenkov waves in a way similar to that is shown in Fig. \ref{fig4}. Likewise, the vortex starts producing V-AV pairs at a critical value $\beta=\beta_s$ that is very close to $\beta_s$ for the laterally infinite stack considered above. The initial stages of the V-AV pair production spreading both along and across the junctions proceeds like it does in the infinite stack, resulting in expanding piles of vortices and antivortices.  However, in the annular JJ stack the propagating macrovortices of opposite polarity eventually collide and partly annihilate as they go through each other. The transient solution then evolves into a chaotically oscillating distribution of $\theta_l(x,t)$ resulting in unidirectional traveling waves of magnetic field with nearly constant amplitudes in each junction, as shown in Fig. \ref{fig10}. Eventually these traveling electromagnetic waves on different layers become more synchronized as shown in Fig. \ref{fig11}. 

Imposing the boundary condition $\theta_1=\theta_N$ models a periodic chain of vortices spaced by $N$ layers along the $z$ direction in an infinite annular JJ stack. Our simulations for this case show that, because of the  symmetry of this geometry, the solutions for $\theta_l(x,t)$ and $B_l(x,t)$ are the same as in the above case of a finite annular stack. 

\section{Vortex chain in an annular stack}
\label{sec:vc}

\begin{figure}
\includegraphics[width=\columnwidth]{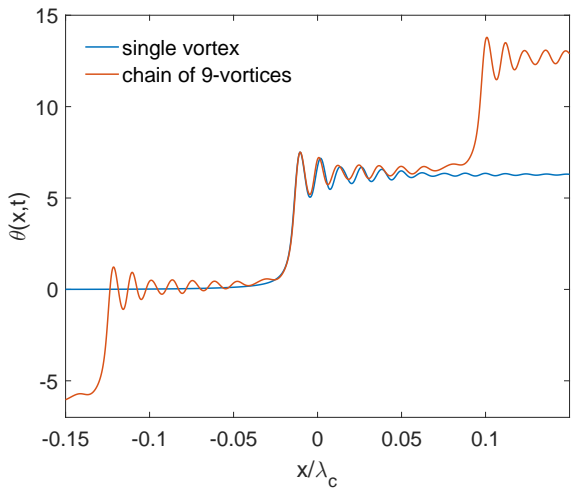}
\caption{Comparison between $\theta_l(x)$ in a single vortex and a chain of 9-vortices (only three are shown) moving along the central junction at $\beta=0.6$.}
\label{fig12}
\end{figure}

The above results show that the initial stage of the continuous V-AV pair production triggered by a single driven vortex is not very sensitive to the boundary conditions either across or along the stack. In this section we present the simulation results for a chain of $M$ vortices placed equidistantly in the middle junction of the 21 JJ stack. If vortices are far apart from each other, so that the spacing between vortices $d=L/M\gg\lambda_J$, the initial stage of the V-AV pair production proceeds in the way similar to that of a single vortex. Namely, each vortex starts radiating Cherenkov wakes at $\beta\approx 0.075$ which matches that of a single vortex for up to $M=9$. The onset of the V-AV pair production at $M=9$ occurs at $\beta=0.625$ close to $\beta_s$ for a single vortex. In this case the intervortex spacing $d\sim 30\lambda_J$ is large so that no significant overlap between the Cherenkov wakes from neighboring vortices happens, as shown in Fig. \ref{fig12}.  

For $M=9$, moving vortices start generating V-AV pairs at $\beta=0.625$. The expanding pairs then overlap, resulting in the phase profile $\theta_{11}(x,t)$ increasing nearly linearly with time while preserving the net winding number of the initial 9 vortices. In turn, the V-AV pair production in the central junction induces V-AV pairs in the neighboring junctions, causing propagation of the resistive state across the stack. Eventually $\theta_l(x,t)$ evolves into a superposition of traveling waves propagating on the phase background increasing linearly with $t$. Our simulations of $M=14$ vortices in the middle layer have shown a similar dynamics of $\theta_l(x,t)$ as for 9 vortices, except that the V-AV pair production starts at a lower value $\beta\approx0.59$. The latter may result from stronger overlap and the constructive interference of  the Cherenkov radiation tails which extend over the length $L_r \sim \lambda_J/\eta$ behind a moving vortex.

\begin{figure}
\includegraphics[width=\columnwidth]{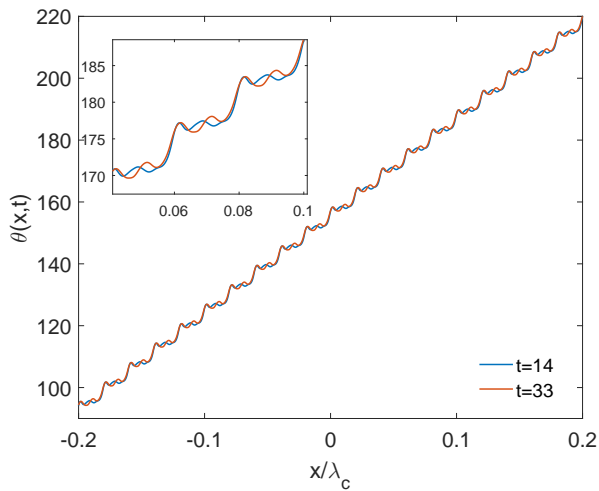}
\caption{Snapshots of $\theta_l(x,t)$ in a moving chain of 50 vortices at $\beta=0.44$ near the instability threshold. The two profiles are superimposed for ease of comparison. Interaction of vortices with Cherenkov wakes causes temporal variations in the shape and velocity of moving vortices.}
\label{fig13}
\end{figure}

\begin{figure}
\includegraphics[width=\columnwidth]{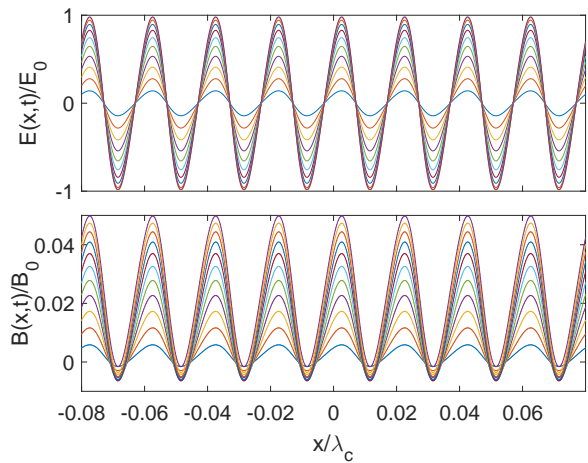}
\caption{Snapshots of the final form of the solution in electric field (top) and magnetic field (bottom) representations in junctions 1-11 for instability current $\beta=0.445$. The oscillations are both in phase and periodic for all layers with amplitudes decaying from the middle junction across the stack.}
\label{fig14}
\end{figure}

The dynamics of vortices changes as the intervortex spacing $d=L/M$ becomes of the order of $\lambda_J$. For instance,  at $M=50$ and $d\simeq 5\lambda_J$ the radiation tails of adjacent vortices overlap even at $\beta\ll \beta_s$. As a result, vortices get trapped in the radiation wakes of neighboring vortices, and the unidirectional motion of the vortex chain at $J$ slightly below $J_s$ is accompanied by a low amplitude traveling wave in which the relative position of the adjacent vortices and their instantaneous velocities oscillate, as shown in Fig. \ref{fig13}. The vortex chain starts producing V-AV pairs at $\beta=0.445$ resulting in a quick  transition of the central junction into a resistive state in which $\theta_{11}(x,t)$ becomes nearly a straight line in $x$ and increases linearly with $t$. Unlike the case of smaller $M$, the quick resistive transition of the central junction does not  spread across the stack and no V-AV pairs are generated on other junctions where only small amplitude plasma traveling waves appear. The electromagnetic oscillations in all layers are phase-locked, the amplitude of oscillations decreasing with the distance from the central layer. Snapshots of these solutions are shown in Fig. \ref{fig14}.

Our simulations have shown that the dynamics of 100 vortices with $d\simeq 2.6\lambda_J$ appears similar to that of 50 vortices. Yet because of stronger overlap of vortices and their Cherenkov radiation tails, the onset of the V-AV pair production $\beta_s=0.455$ is slightly higher than for 50 vortices.  This trend becomes more apparent for 200 vortices for which $\beta_s\simeq 0.665$ not only exceeds $\beta_s$ for 100 vortices but also $\beta_s$ for a single vortex.  The increase of $\beta_s$ with $M$ at large $M$ may result from the fact that, if vortices and their radiation tails overlap strongly, the spatial modulations of $\theta(x,t)$ along the vortex chain get reduced, and the critical $\pi$ phase nucleus which triggers the V-AV pair production can only appear at higher $\beta$. For a very dense vortex chain with $d\ll \lambda_J$, the V-AV pair production does not occur before the central junction switches to the resistive state at $\beta=1$.   

\section{vortex lattice}
\label{sec:vl}

In this section we present the results of our simulations for the driven Josephson vortex lattice in an annular stack of planar junctions.

\subsection{Annular stack with finite $N$}

\begin{figure}
\includegraphics[width=\columnwidth]{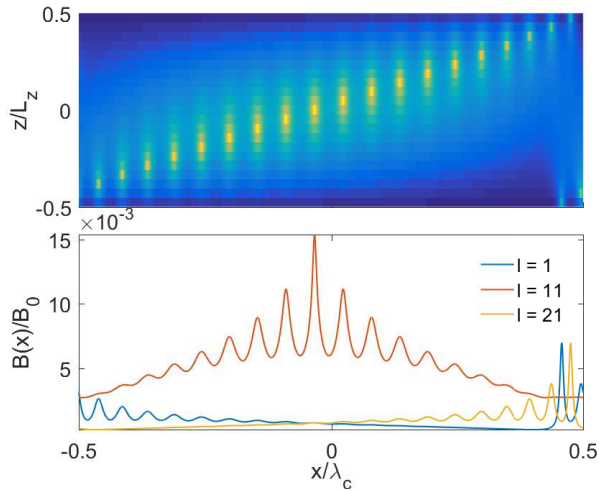}
\caption{Color map of the magnetic field across the stack for a stationary vortex lattice with one fluxon per layer (top) and $B_l(x)$ for the middle and surface JJs (bottom). }
\label{fig15}
\end{figure}

\begin{figure}
\includegraphics[width=\columnwidth]{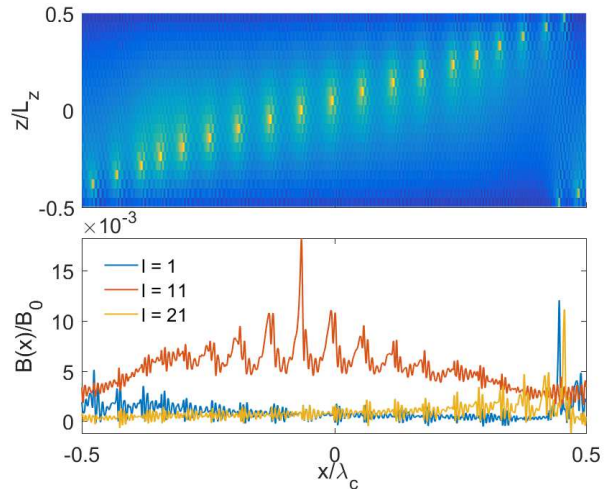}
\caption{Color map of the magnetic field across the stack for a uniformly moving vortex lattice with one fluxon per layer (top) and $B_l(x)$ for the middle and surface JJs (bottom) calculated at $\beta=0.54$.}
\label{fig16}
\end{figure}

Consider vortices initially placed along a line slightly tilted from being perpendicular to the layers with one vortex per layer in an annular stack with $N=21$. At zero current this structure then relaxes to that is shown in the top panel of Fig. \ref{fig15}. The corresponding field distributions $B_l(x)$ are shown in the bottom panel of Fig. \ref{fig15} for the top most, bottom most and middle layer. After a bias current is applied the vortices start moving uniformly and radiating Cherenkov waves with the amplitude and wavelengths increasing with $\beta$. As shown in Fig. \ref{fig16}, the average velocities of vortices in different layers are almost the same and their relative positions remain constant as the current is ramped up to the onset of the V-AV pair production, $\beta=0.54$. At $\beta_s=0.55$ the vortex moving with the velocity $v\approx 1.34c_s$ along the $20$-th junction starts generating V-AV pairs which then spread to other junctions, driving the whole stack into a resistive state. As a result, the initial vortex structure evolves to $\theta_l(x,t)$ which appears chaotic in both $x$ and $t$ on each junction, similar to that was obtained for a single vortex shown in Fig. \ref{fig10}.

\begin{figure}
\includegraphics[width=\columnwidth]{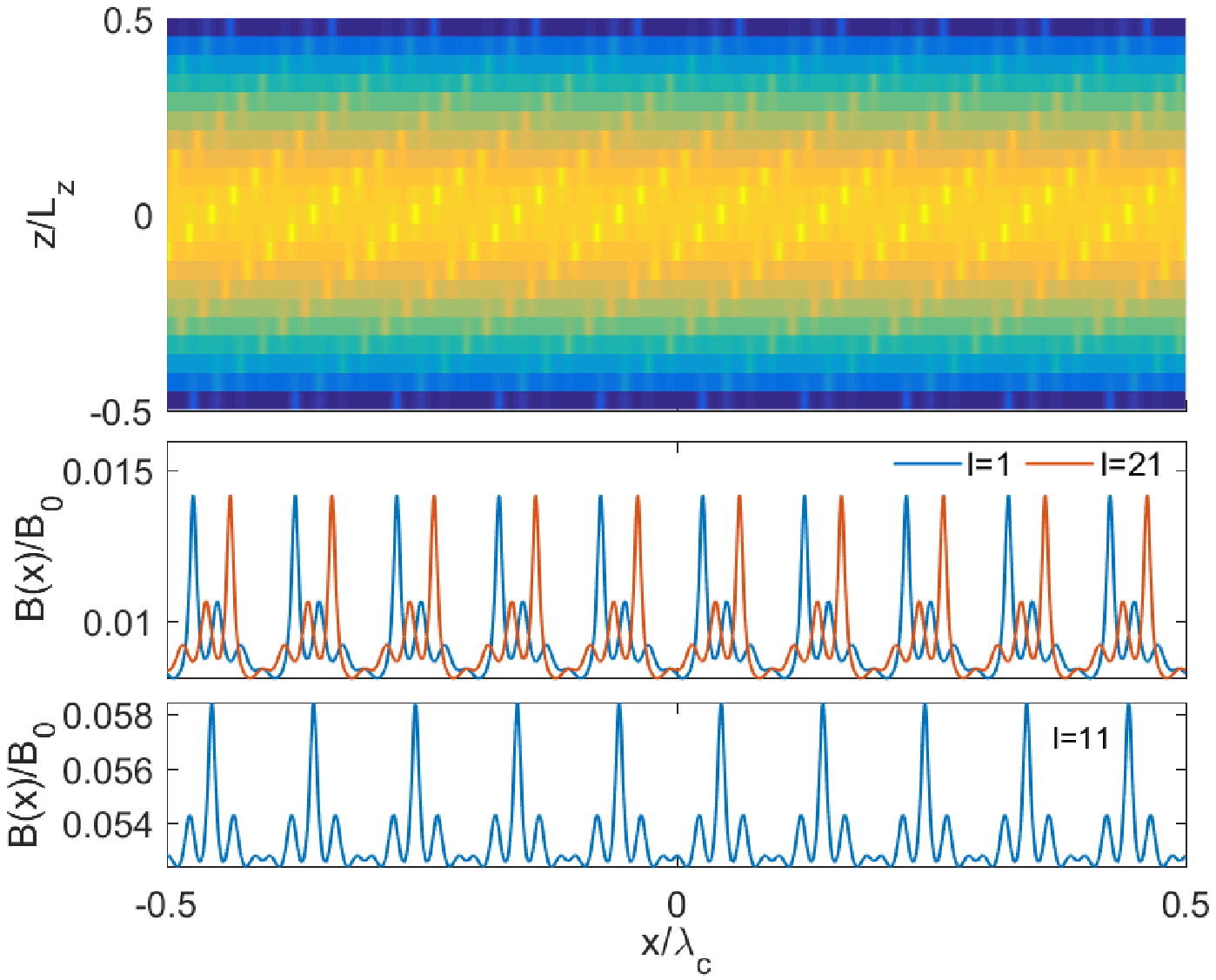}
\caption{Color map of the magnetic field across the stack for a stationary vortex lattice with ten fluxon per layer (top) and $B_l(x)$ for the middle and surface JJs (bottom). }
\label{fig17}
\end{figure}

\begin{figure}
\includegraphics[width=\columnwidth]{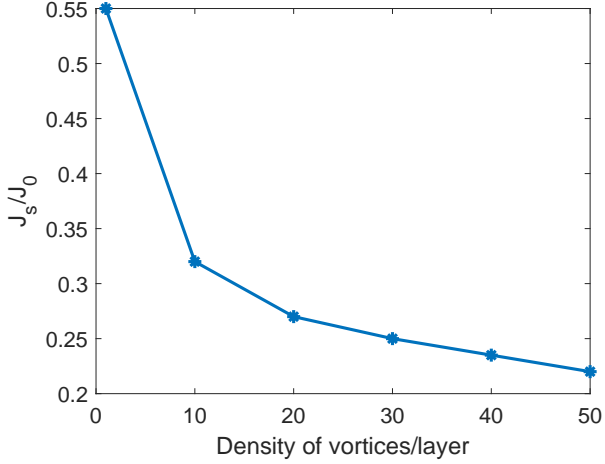}
\caption{Calculated dependence of $J_s$ on the linear density of vortices per length $\lambda_c$ along the layer in a vortex lattice. }
\label{fig18}
\end{figure}

In our numerical simulations we observed that the symmetry of static vortex structures can depend strongly on the initial arrangement of vortices which can relax to many metastable states. This issue has been recognized in the literature as one of the main reasons why vortices do not necessarily form a triangular lattice in numerical simulations \cite{jvlins,insrecjvl}. To produce a static vortex configuration with equidistant arrangement of vortices, we initially put chains of equidistant vortices in each layer with vortices on neighboring layers shifted with respect to each other. As a result, vortices relax to a periodic structure, as shown in Fig. \ref{fig17} for ten vortices per layer. We found that, for a current-driven vortex lattice, the onset of the V-AV pair production is mostly determined by the vortex density within each layer and depends weakly on the symmetry of the vortex lattice.  For instance, for the structure shown in Fig. \ref{fig17}, the V-AV pair production occurs at $\beta\approx 0.32$ irrespective of the arrangement of vortices as long as the linear density of vortices per junction is fixed.  From our calculations, it follows that the threshold current $J_s$ decreases monotonically with the increase of the linear density vortices per layer as shown in Fig. \ref{fig18}. Hence, $J_s$ is reduced if a weak parallel magnetic field is applied to the stack.  

\begin{figure}
\includegraphics[width=\columnwidth]{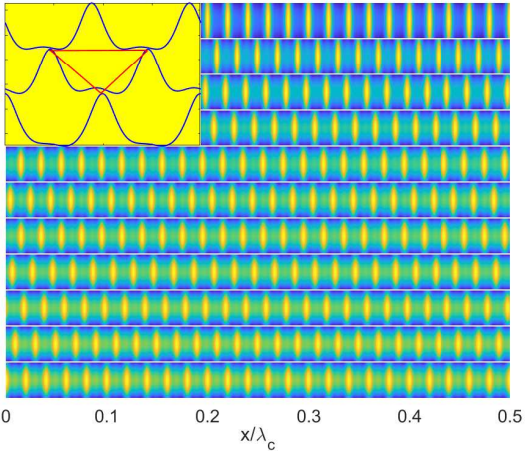}
\caption{Color map of the magnetic field in a stationary vortex lattice composed of fifty fluxons per layer. The close up in the top left corner shows a triangle formed by three vortices in two adjacent layers. }
\label{fig19}
\end{figure}

As the density of vortices is increased the vortex configuration becomes closer to a triangular lattice, as shown in Fig. \ref{fig19} for a lattice of $50$ vortices per layer. If a bias current is applied, Cherenkov radiation occurs once the velocity of the lattice exceeds the threshold for the minimum plasma mode, but the radiation wakes are reduced due to strong overlap of vortices in both directions. Here the chain of vortices in the middle junction become unstable first at $\beta_s=0.195$ producing only one V-AV pair after which the pair production stops. At a slightly larger current of $\beta=0.2$ two more V-AV pairs are generated in the neighboring 10-th and 12-th junctions, while larger number of V-AV pairs are produced in the middle junction. As current is increased to $\beta=0.205$ some vortices in the 9-th and 13-th junctions produce a few V-AV pairs. This stepwise process of limited V-AV pair production spreads across more and more junctions as the current further increases.  Finally, at $\beta=0.2225$ the middle junction starts generating V-AV pairs, which triggers the V-AV pair production in all JJs. As a result, at $\beta>0.2225$ the stack eventually switches into a dynamic resistive state comprised of propagating phase-locked waves which are synchronized for all junctions.   

\subsection{Annular stack with $\theta_1=\theta_N$}

\begin{figure}
\includegraphics[width=\columnwidth]{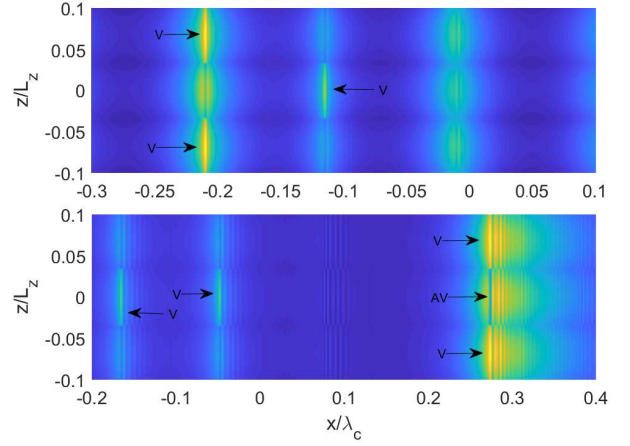}
\caption{Magnetic field color map in moving vortices in junctions $10$, $11$ and $12$ at $\beta=0.1$ (top). Bottom panel illustrates how a transient triplet is formed out of the conversion of the image of vortices from 10th and 12th junction in the central JJ to a pair of V-AV at $\beta=0.175$.}
\label{fig20}
\end{figure}

Here we impose the periodic boundary condition of $\theta_1=\theta_{N}$ which model periodic vortex structures in an annular stack infinite along $z$. Due to the symmetry of the problem, this boundary condition reduces the number of variables $\theta_l(x,t)$ in Eqs. (\ref{csg}) to $(N+1)/2$ for odd $N$. Consider one fluxon per layer for which the situation is similar to that considered in the previous section. Bcause of the exact same position of vortices in 10-th and 12-th junctions, the magnitude of the image induced by these vortices on the middle junction $(l=11)$ doubles. As a result, the onset of the V-AV pair production on the central junction is reduced down to $\beta_s=0.175$. At $\beta=\beta_s$ this image in the middle junction converts to a V-AV pair which then expand in such a way that two vortices move to the left and the antivortex moves to the right until it gets trapped between two vortices in the neighboring junctions 10 and 12. Shown in Fig. \ref{fig20} are snapshots of magnetic field maps at $\beta<\beta_s$ and $\beta>\beta_s$ which illustrate the formation of transient V-AV-V triplets.  The antivortex trapped in the V-AV-V triplet slows it down relative to other vortices, so when the vortices from junction 9 and 13 reach the triplet, the antivortex escapes, producing a V-AV pair which then annihilates, as shown in the simulation movie \cite{supp}. The process of creation and then annihilation of pairs during the disintegraion of the triplet occurs as $\beta$ further increases. Finally, at $\beta=0.3$ after the disintegration of the triplet, a cascade of V-AV pairs generated  continuously in the central junction spreads across the whole stack, resulting in a McCumber-type resistive state in which $\theta_l(t)$ on each junction increases nearly linear with time \cite{supp}. 

\section{Finite size effects and vortex bouncing}
\label{sec:finsize}

Proliferation of branching V-AV patterns or macrovortex (MV) structures caused by a single vortex is essentially a bulk effect which occurs in a sufficiently long sample or an annular JJ stack.  However, in a JJ stack of finite length $L_x$, the expanding MV chain eventually hits the edges of the JJs, where the boundary conditions of zero current $\theta_l^\prime=0$ are imposed.  In this section we consider peculiarities of vortex dynamics resulting from the finite size effects. It turns out that interaction of a MV with the edges of the stack occurs in a way similar to that of a moving J vortex in a single long JJ (see, e.g., Ref. \onlinecite{physc}). This interaction proceeds as follows. As V approaches the edge of a JJ, it induces penetration of a counter-propagating AV which collides with the incoming vortex. The outcome of this collision depends on the damping parameter $\eta$.  In an overdamped JJ  $(\eta \gtrsim 1)$, the colliding V and AV annihilate, fully extinguishing the fluxon of the initial vortex as it exits the junction. However in an underdamped JJ with $\eta\ll 1$, the colliding V and AV do not annihilate but go through each other, as characteristic of non-dissipative solitons described by the sine-Gordon equation \cite{BP}. As a result, the incoming V exits while the AV moves into the JJ. This process can be regarded as a vortex analog of the Andreev reflection.  

\begin{figure}
\includegraphics[width=\columnwidth]{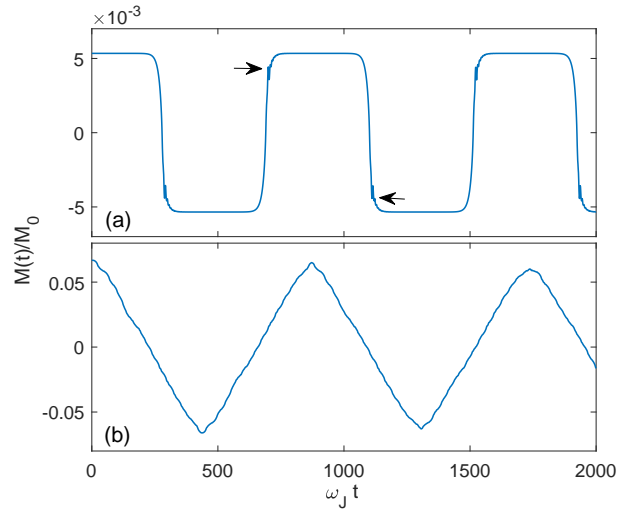}
\caption{Temporal oscillations of a magnetic moment $M(t)$ due to periodic reflections of driven vortices and antivortices from the sample edges at $\eta=0.1$. (a) $M(t)$ caused by a vortex shuttle in which a single vortex gets reflected from the edges as antivortex at $\beta=0.585<\beta_s$. The features marked by the arrows result from Cherenkov and bremstrahlung radiation after reflection of a V or AV. (b) $M(t)$ caused by a bouncing flux structure with one vortex per layer at $\beta=0.53<\beta_s$. }
\label{fig21}
\end{figure}

A current-driven V in an underdamped JJ stack gets periodically reflected from the edge where it transforms into a counter-propagating AV which in turn gets reflected as a vortex from the opposite edge. Such V-AV shuttle causes oscillations of the magnetic moment $M(t)$ with the flight frequency $\nu=v/2L_x$ depending on the JJ length. Here $M(t)=\phi(t)L_y$ and the instantaneous magnetic flux threading the stack $\phi(t)$ are calculated using
\begin{equation}
M(t)=M_0\sum_l\int_0^{L_x}B_l(x)dx,
\label{M}
\end{equation} 
where $M_0=B_0 s\lambda_c L_y/\mu_0=\phi_0L_y/2\pi\mu_0$, $L_y$ is the length of the stack along $y$, and the integral is expressed in terms of the dimensionless field $B_l$ and coordinates defined in Sec. \ref{sec:elec}.
Shown in Fig. \ref{fig21}a is $M(t)$ calculated for a vortex driven along the central layer at $\beta <\beta_s$ in a stack with $L_x=\lambda_c$ and $N=21$. The magnitude of $|M(t)|\simeq 0.0055 M_0$ in Fig. \ref{fig21}a indicates that the vortex flux $\phi\simeq 9\cdot 10^{-4}\phi_0$ is much smaller than $\phi_0$. This effect is similar to the well-known reduction of magnetic flux in a parallel Abrikosov vortex in a thin film \cite{vf1,vf2,vf3}. Calculation of $\phi$ of a vortex in a long JJ stack with $N\gg 1$ and $L_x\gg \lambda_J$ given in Appendix \ref{Ap} yields the same result as for the Abrikosov vortex \cite{vf2}:   
\begin{equation}
\phi(u) = \phi_0\left[1-\frac{\cosh(u/\lambda_{ab})}{\cosh(L_z/2\lambda_{ab})}\right].
\label{ph}
\end{equation} 
Here $u$ is the position of the vortex relative to the center of the film. Notice that $\phi(u)$ decreases as $u$ increases and vanishes at the surface $u=\pm L_z/2$ where the vortex flux is extinguished by AV images \cite{vf1,vf2}. For the J vortex in the center of a thin JJ stack $(u=0,\, L_z=sN\ll 2\lambda_{ab})$, Eq. (\ref{ph}) gives:
\begin{equation}
\phi\simeq \frac{\phi_0 N^2}{8}\left(\frac{s}{\lambda_{ab}}\right)^2,\qquad N\lesssim \frac{2\lambda_{ab}}{s}.
\label{phi}
\end{equation}
Taking here $N=21$, $s=1.5$ nm and $\lambda_{ab}=400$ nm for BSCCO, we obtain $\phi\simeq 8\cdot 10^{-4}\phi_0$ in agreement with the simulation results presented in Fig. \ref{fig21}a.

Shown in Fig. \ref{fig21}b is $M(t)$ calculated for a dynamic flux state with one vortex per layer below the Cherenkov instability threshold at $\beta <\beta_s$.  Here the magnitude of $M(t)$ for 21 vortices is about 12 times larger than for a single vortex. The fact that $M(t)$ for one vortex per layer is not 21 times larger than $M(t)$ for a single vortex is consistent with Eq. (\ref{ph}) according to which the flux of vortices on outer layers is smaller than $\phi$ for the vortex on the central layer. The shape of $M(t)$ changes from rectangular pulses for a single vortex to triangular pulses for many vortices. This happens because the repelling vortices tend to arrange themselves to maximize the intervortex spacing so the reflections of vortices from the edges on different layers occur at different times.        

\begin{figure}
\includegraphics[width=\columnwidth]{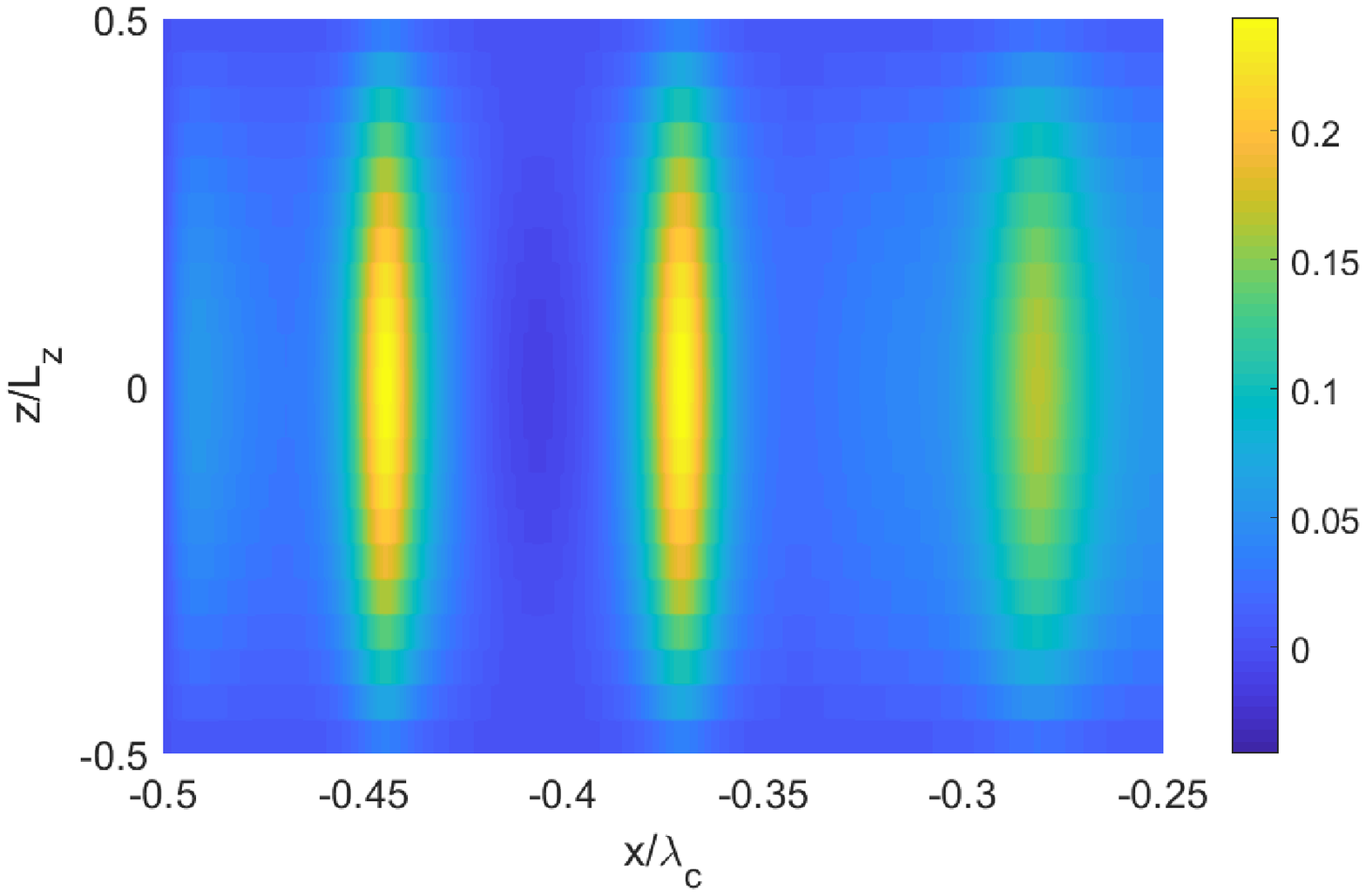}
\includegraphics[width=\columnwidth]{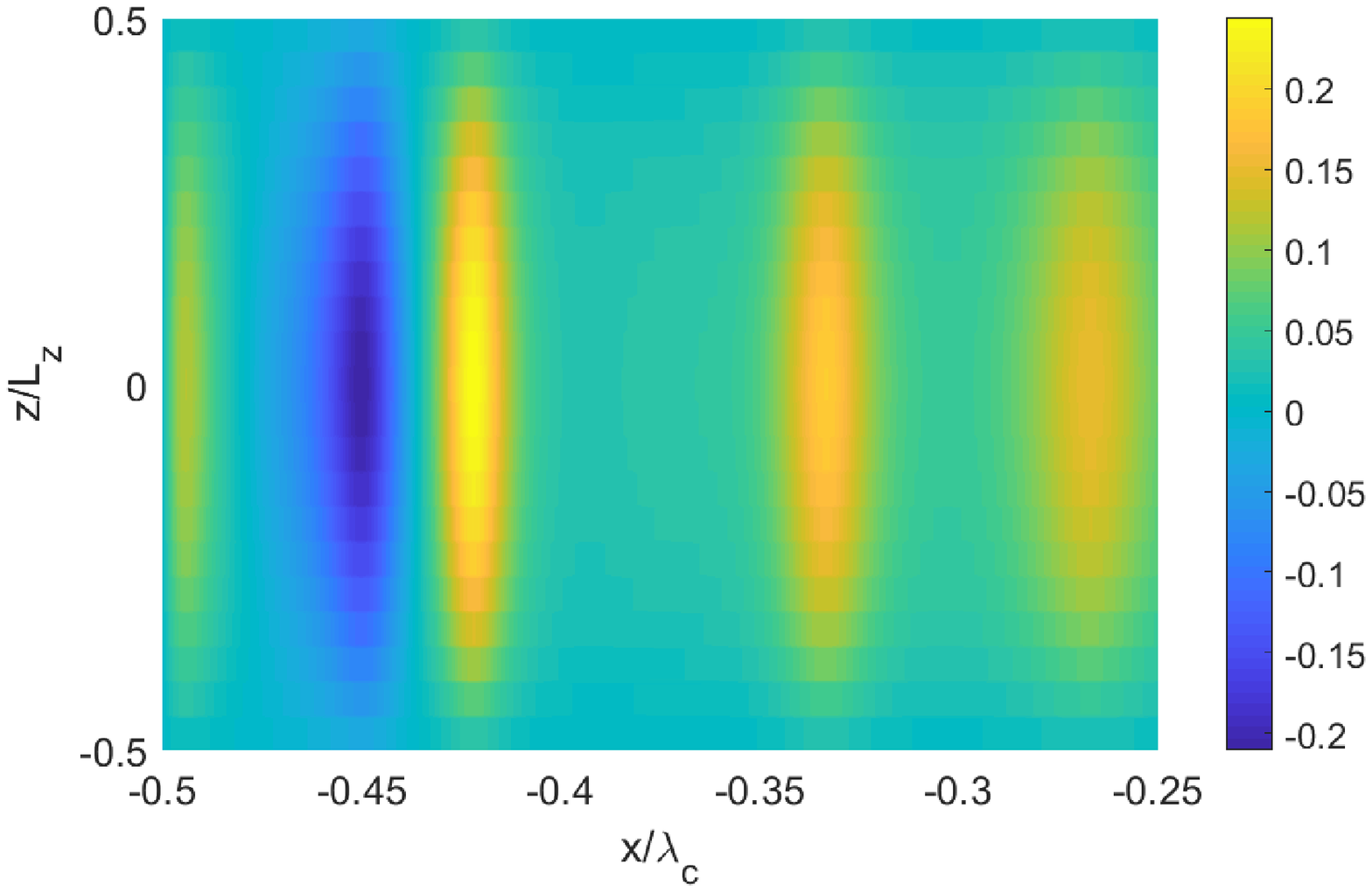}
\caption{Magnetic field color map in moving macrovortices colliding with the edge of the stack at $x/\lambda_c=-0.5$.  Top: A chain of macrovortices reaching the edge just before the collision. Bottom: The same chain after the leading macrovortex collided with the edge and got transformed into a counter-propagating anti-macrovortex. }
\label{fig22}
\end{figure}

\begin{figure}
\includegraphics[width=\columnwidth]{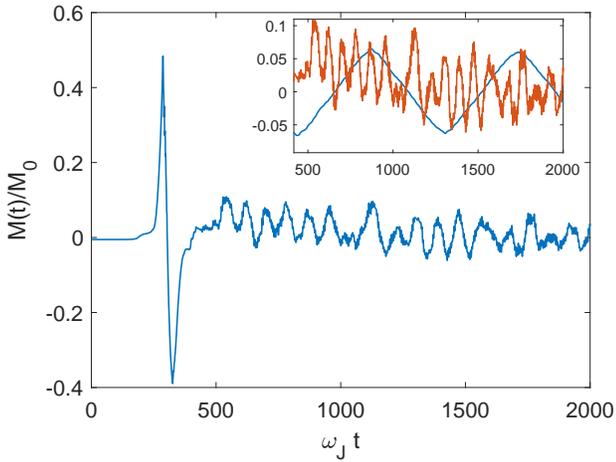}
\caption{Temporal magnetic moment $M(t)$ due to bouncing macrovortices excited by a single V-AV shuttle. Inset shows $M(t)$ caused by self-sustained MV standing waves  superimposed onto $M(t)$ due to stable oscillations of the flux structure with one vortex per layer taken from Fig. \ref{fig21}.}
\label{fig23}
\end{figure}

Above the Cherenkov instability threshold $\beta>\beta_s$ a single V-AV shuttle excites counter-propagating MVs and anti-macrovortices (AMV) which then get reflected from the edges in the same way as single Vs and AVs. For instance, the collision of MVs with the edge of an underdamped stack with $\eta=0.1$ is shown in Fig. \ref{fig22}. As the MV exits the stack it induces penetration of a counterpropagating AMV, the structure of this AMV remains preserved as it goes through the incoming MV without fragmentation into single vortices. Such bouncing MVs and AMVs generated by a V-AV shuttle give rise to temporal oscillations of the magnetic moment $M(t)=L_y\phi(t)/\mu_0$, where $\phi(t)$ is the net magnetic flux produced by all Vs and AVs. As shown in Fig. \ref{fig23}, the magnitude of $M(t)$ is of the order of that of a stable flux structure with one vortex per layer (see Fig. \ref{fig21}b). Notice that $M(t)$ for bouncing MVs contains multiple harmonics with frequencies much higher than those for the stable flux structures shown in Fig. \ref{fig21}. 

\begin{figure}
\includegraphics[width=\columnwidth]{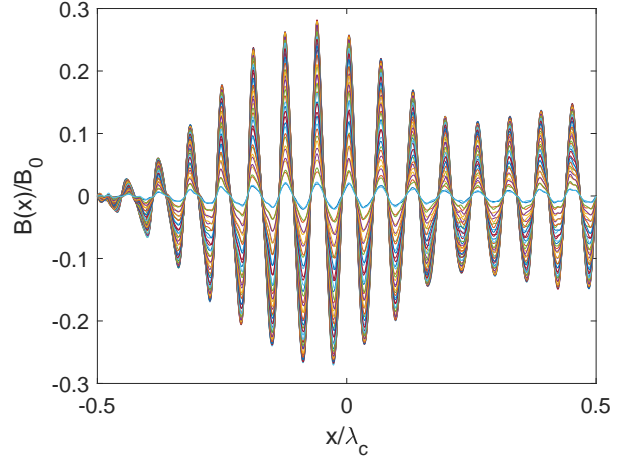}
\caption{A snapshot of beating standing waves of $B_l(x,t)$ on different layers in a finite stack with $N=41$ calculated for self-sustained oscillations of $M(t)$ shown in Fig. \ref{fig23}.}
\label{fig24}
\end{figure}

A big transient spike in $M(t)$ at the onset of the MV formation can be understood as follows. At $\beta>\beta_s$ the initial vortex placed near the right edge of the stack accelerates and starts producing V-AV pairs which form the MV structures spreading both along and across the JJ stack. Here MVs move to the left along with the initial vortex while AMVs move to the right and get reflected as MVs from the right edge before the leading MV reaches the left edge. As a result, the number of vortices in the stack keeps growing until the leading MV reaches the left edge, after which the process reverses as the number of AMVs increases and exceeds the number of MVs. After a few bouncing of MVs and AMVs back and forth, generation of new V-AV pairs stops and a standing wave, resulting in self-sustained oscillations of $M(t)$ forms, as shown in Fig. \ref{fig23}. A snapshot of this standing wave in Fig. \ref{fig24} indicates nonlinear interference and multiplication of harmonics with frequencies ranging from $\omega\sim\omega_J$ to much lower frequencies $\omega \sim v/d$ determined by the velocity $v(\beta)$ and the spacing $d(\beta)$ between MVs. Simulation movies of this process are available in Ref. \onlinecite{supp}.
 
Self-sustained MV standing waves excited by a V-AV shuttle at $J>J_s$ increase the power of electromagnetic radiation $W$ caused by temporal oscillations of $M(t)$ and a charge density at the surface of the stack. We do not consider here all essential contributions to $W$ which depend on the geometry of the stack and details of its electromagnetic coupling with surrounding structures (see, e.g., reviews \onlinecite{thz1,thz2,lin-sust} and the references therein) but only estimate a magneto-dipole part of  $W$ which has not been addressed in the literature.  As follows from the inset in Fig. \ref{fig23}, each MV at $N=21$ has $\sim N\phi_0$ bunched vortices lined perpendicular to the layers. Such bouncing multi-quanta MVs greatly increase the magneto-dipole radiation power $W\propto \ddot{M}^2$ as compared to the V-AV shuttle at $\beta<\beta_s$. Indeed, once $J$ exceeds $J_s$, both the magnitude and the frequency of $M(t)$ shown in Figs. \ref{fig21} and \ref{fig23} increases by more than an order of magnitude, which translates to $\sim 10^7$ fold increase in $W$. 

\begin{figure}
\includegraphics[width=\columnwidth]{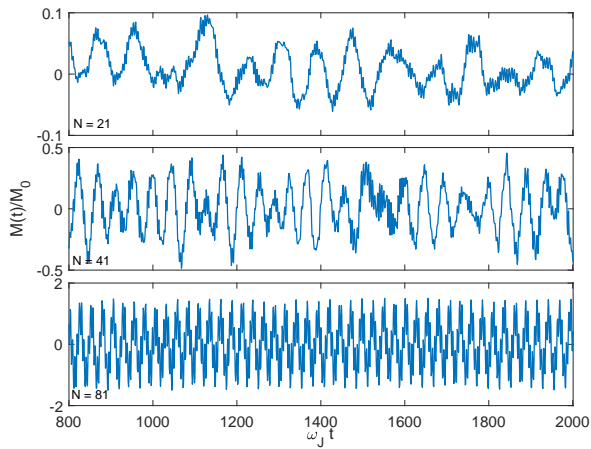}
\caption{Self-sustained oscillations of $M(t)$ calculated for $N=21$, $N=41$ and $N=81$ at $\beta=0.6$ and $\eta=0.1$ after complete decay of initial transient spikes in $M(t)$.}
\label{fig25}
\end{figure}

\begin{figure}
\includegraphics[width=\columnwidth]{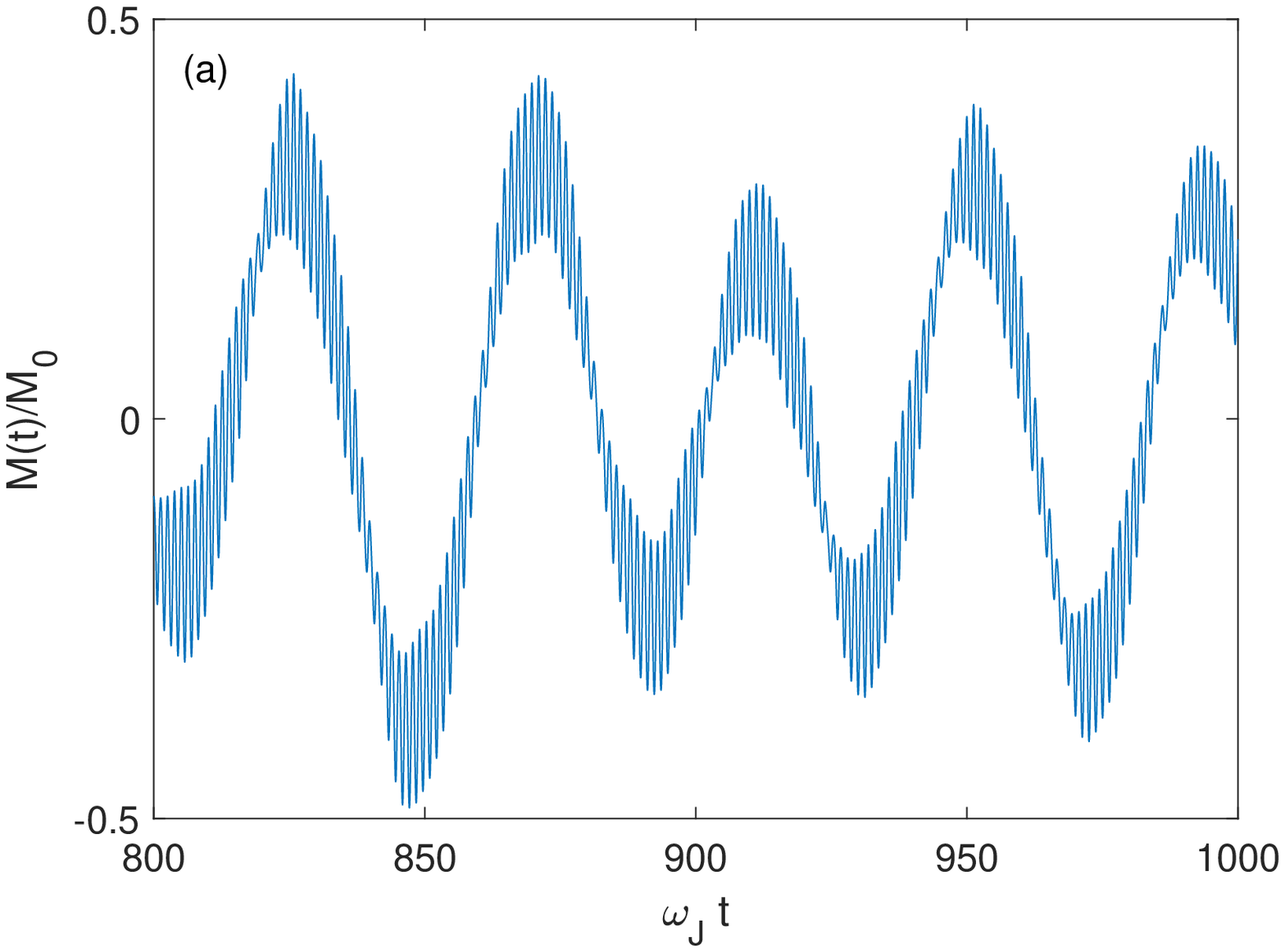}
\includegraphics[width=\columnwidth]{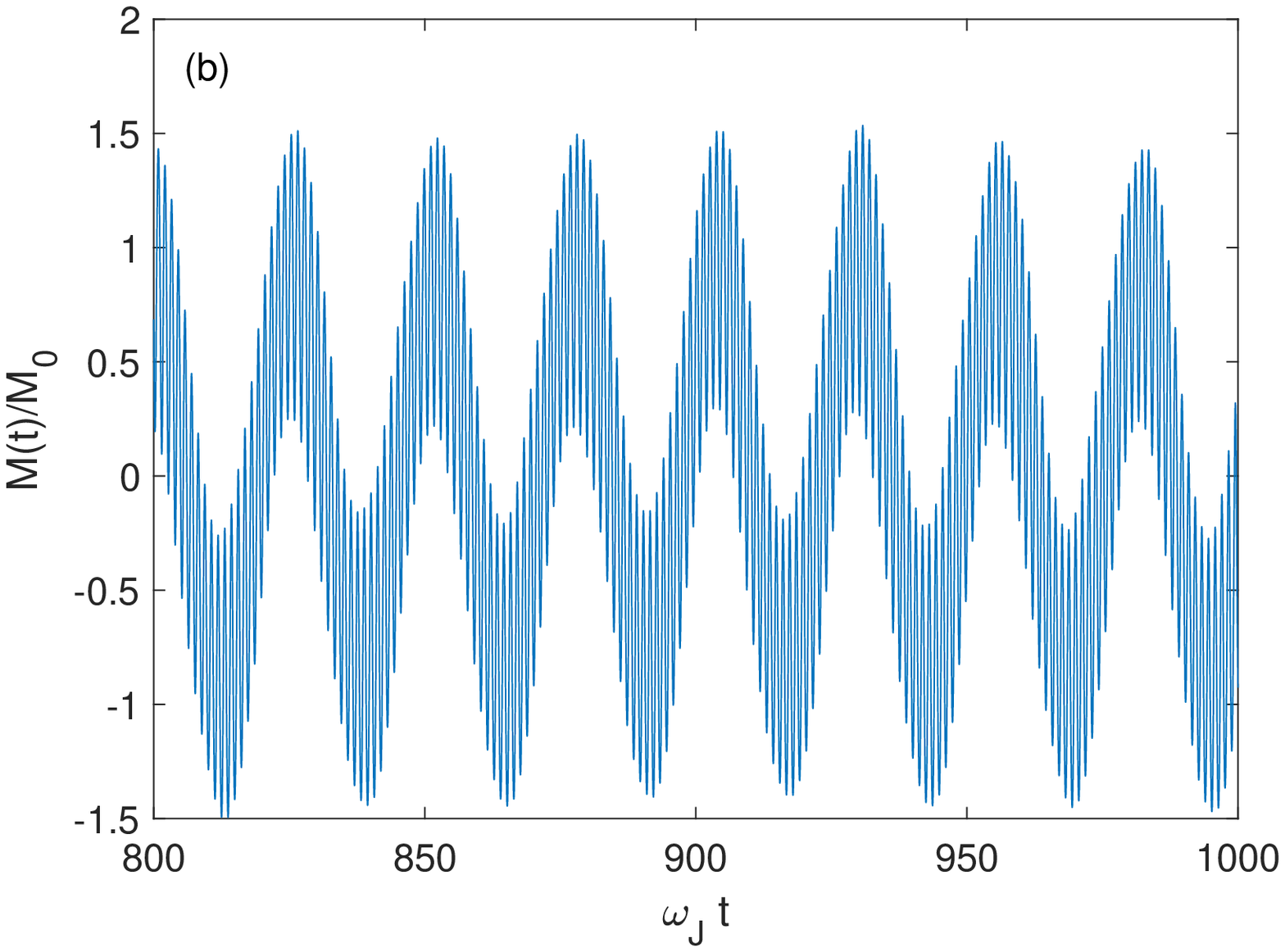}
\caption{Parts of $M(t)$ at $N=41$ and $N=81$ shown in Fig. \ref{fig25} but calculated with the finer time steps $\Delta t =0.01\omega_J^{-1}$ 
to reveal high-frequency harmonics in $M(t)$.}
\label{fig26}
\end{figure}

Both the magnitudes and the frequencies of different harmonics in $M(t)$ change significantly as the number of layers increases. Shown in Fig. \ref{fig25} are $M(t)=\phi(t)L_y$ calculated at $N=21$, $N=41$, and $N=81$ after the transient spikes in $M(t)$ decayed completely. Parts of these $M(t)$ curves calculated with much finer time steps $\Delta t=0.01\omega_J^{-1}$ shown in Fig. \ref{fig26} clearly exhibit multiple harmonics with high frequencies $\omega\sim \omega_J$ and low beating frequencies $\omega\ll \omega_J$ which increase nearly linearly with $N$. As was mentioned above, the low-frequency part of $M(t)$ is related to traveling times of MVs. Characteristic magnitudes $M_N$ of $M(t)$ also increase as $N$ increases: $M_{81}\simeq 4M_{41}$ and $M_{41}\simeq (4-5) M_{21}$. This trend is qualitatively consistent with the quadratic increase of the  magnetic flux per vortex $M_N \propto \phi \propto N^2$ in $JJ$ stacks with $L_z\ll 2\lambda_{ab}$ given by Eq. (\ref{phi}).

The mean radiation power $W=\mu_0\langle \ddot{M}^2\rangle/6\pi c^3$ for JJ stacks smaller than the radiated wavelength \cite{griffiths} can be estimated using $M(t)$ from Eq. (\ref{M}), where   $M_0=\phi_0L_y/2\pi\mu_0$ and $\omega_J=c/\sqrt{\epsilon_c}\lambda_c$. Hence, $W$ can be presented in the form  
\begin{equation}
W\simeq \frac{c(\phi_0L_y)^2G_N}{24\pi^3\mu_0\epsilon_c^2\lambda_c^4},
\qquad 
G_N=\int_{t_0}^{t_0+T}\ddot{m}^2\frac{dt}{T},
\label{ww}
\end{equation}
where $m(t)=M(t)/M_0$. The dimensionless factor $G_N$ takes into account the effect of the number of layers on the amplitudes and frequencies of different harmonics in $M$ which contribute to $W$, where $t_0\simeq 800$. We evaluated $G_N$ by averaging numerical derivatives in $\ddot{m}^2$ for the calculated $M(t)$ over the time interval $T=200$. Calculations of $G_N$ for different $N$ using the results shown in Fig. \ref{fig26} give $G_{21}=0.0336$, $G_{41}=2.05$ and $G_{81}=154.1$. Such strong increase of $G_N$ with $N$ is much faster than $W\propto N^4$ resulting from only the quadratic increase of the magnetic flux of the vortex with $N$. Another part of this rapid growth of $G_N$  comes from the enhancement of higher-frequency harmonics at larger $N$ evident from Figs. \ref{fig25} and \ref{fig26}.  All in all, the calculated $G_N$ roughly follows the $N^6$ dependence at $N\lesssim 10^2$. 
 
Taking $\lambda_c=200\, \mu$m,  $L_y=1$ mm, $\epsilon_c=10$, and $G_{81}=154$ in Eq. (\ref{ww}), we obtain $W\simeq 1.32$ nW of the order of the lower end of radiated power observed on BSCCO mesas \cite{thz2,thz3} with a much larger number of layers $N\sim 10^3$. Yet given the very rapid increase of $W_N\propto N^6$ revealed in our simulations at $N\lesssim 10^2$, a much greater $W$ at $N\sim 10^3$ may occur. Direct calculation of $W$ for $N\sim 10^3$ is beyond our current computational capabilities. Yet if the trend $W\propto N^6$ would continue up to $N\simeq 2\lambda_{ab}/s\simeq 500$ at which the flux per vortex reaches $\phi_0$ (see Eqs. (\ref{ph}) and (\ref{phi})), one might expect $W_{500}\sim W_{81}(500/81)^6 \sim1$ mW (for an ideal cooling of the sample and no Joule heating caused by the motion of MVs).    

\section{Discussion}
\label{sec:disc}

In this paper we show that uniform motion of a Josephson vortex driven by a dc current in layered superconductors breaks down as the velocity of the vortex exceeds the terminal velocity $v_c$ at current densities $J>J_s$. If  $v>v_c$ the moving vortex starts emitting V-AV pairs, causing a dendritic flux branching in which vortices and antivortices become spatially separated and form dissipative structures which depend on the sample geometry. For instance, a single vortex in a long stack can produce a chain of dissipative macrovortices that extend across the entire stack as shown in Fig. \ref{fig9}. The breakdown of the dc flux flow state caused by V-AV pair production can occur at current densities $J_s$ well below the Josephson critical currents $J_0$ across the stack. 

In an underdamped JJ stack of finite length $L_x$ a vortex driven by a dc current at $J<J_s$ turns into a V-AV shuttle in which the vortex periodically changes its polarity and direction of motion after each reflection from the sample edge.  This process results in oscillations of the magnetic moment $M(t)$ with the flight frequency $v/2L_x$ depending on the length of the stack. At $J>J_s$ the V-AV shuttle produces propagating macrovortices consisting of bunched vortices aligned perpendicular to the layers. These macrovortices periodically change both the polarity and the direction of motion without fragmentation into single vortices after each reflection from the edges of the JJ stack. Such bouncing macrovortices eventually form large-amplitude flux standing waves, giving rise to oscillations of $M(t)$. Here $M(t)$ contains multiple harmonics the amplitudes and frequencies of which increase as the number of layers increases.      

Proliferation of V-AV pairs at $J>J_s$ can manifest itself in hysteretic jumps on the V-I curves. These jumps appear similar to those produced by heating effects\cite{KL,thz2} yet the initial stage of the Cherenkov vortex instability is affected
by neither cooling conditions nor the nonequilibrium kinetics of quasiparticles. Moreover, heating is most pronounced in overdamped junctions with $\eta>1$ in which radiation is suppressed, whereas the Cherenkov instability is most pronounced in weakly-dissipative underdamped interlayer junctions characteristic of the BSCCO cuprates.  The V-AV pair production can be facilitated by interaction of vortices with edges or materials defects, resulting in vortex bremsstrahlung and further reduction of the terminal velocity $v_c$ and the threshold of instability current density $J_s$. These effects are similar to those revealed in our previous simulations of current-driven vortices in a single Josephson junction of finite length \cite{screp}.   
  
The V-AV pair production and bouncing macrovortices caused by a single vortex at $J>J_s$ can contribute to the power of radiation $W$ from a JJ stack.  As was shown in Sec. \ref{sec:finsize}, the V-AV shuttle generates self-sustained MV standing waves and oscillations of the total magnetic moment. In turn, oscillations of $M(t)$ gives a contribution to the radiation power which increases greatly as the number of layers increases. For the parameters of BSCCO and $N\leq 81$ our calculations gave $W\sim 1$ nW, so one might expect $W\sim 1$ mW at $N\sim 10^3$ characteristic of the BSCCO mesas. Hence,  bouncing macrovortices could contribute to the radiation power observed in the BSCCO mesas, although specifying the fraction of this contribution in the total $W$ requires more elaborate calculations taking into account the sample geometry and cooling conditions.  The nonlinear MV standing wave at $J>J_s$ eventually give rise to strong dissipation which can produce hotspots in the sample \cite{hs1,hs2}, even though heating is not the underlying cause for the V-AV pair production but rather its consequence. Our results thus suggest a mechanism by which the formation of hotspots may be linked to peaks in the radiation power, as was indeed observed on the BSCCO mesas \cite{ths1,ths2,ths3,ths4,ths5}. 

\section*{Acknowledgments}

This work was supported by the US Department of Energy under Grant No. DE-SC0010081-020. We thank A.E. Koshelev for a useful discussion.

\appendix

\section{Magnetic flux of a parallel J vortex} \label{Ap}

We calculate the magnetic flux $\phi$ of a vortex in a long JJ stack with $N\gg 1$ and $L_x\gg \lambda_J$. 
The vortex core has the length $\lambda_J=s\Gamma$ along the layer and a width $\sim s$ across the layers. At $\Gamma\gg 1$ the magnetic field varies slowly across the neighboring layers, so the discrete $B_l(x)$ can be approximated by a continuous function $B(x,y)$ which satisfies the anisotropic London equation:
\begin{equation}
\lambda_{ab}^2\frac{\partial^2 B}{\partial z^2}+\lambda_{c}^2\frac{\partial^2 B}{\partial x^2} - B =-\frac{\phi_0}{2\pi}\frac{\partial \varphi}{\partial x}\delta(z-u),
\label{a1}
\end{equation}  
where $\varphi(x)$ is a $2\pi$ kink of length $\lambda_J$ which describes the phase difference between the layers where the vortex core is located at $z=u$. 
The boundary conditions of zero current through the surface requires $B(x,\pm L_z/2)=0$.

The magnetic flux is given by 
\begin{equation}
\phi=\int_{-\infty}^{\infty}dx\int_{-L_z/2}^{L_z/2}B(x,z)dz=\int_{-L_z/2}^{L_z/2}g(z)dz,
\label{a3}
\end{equation}
where $g(z)=\int_{-\infty}^{\infty} B(x,z)dx$, and $z=0$ is taken in the center of the stack. The equation for $g(z)$ is obtained by 
integrating Eq. (\ref{a1}) over $x$ from $-\infty$ to $\infty$, using the boundary conditions $\partial_xB(\pm\infty,z)=0$ and $\varphi(\infty)-\varphi(-\infty)=2\pi$. 
Hence,
\begin{equation}
\lambda_{ab}^2\frac{\partial^2 g}{\partial z^2} - g =-\phi_0\delta(z-u).
\label{a4}
\end{equation}
The solution of Eq. (\ref{a4}) satisfying the boundary condition $g(\pm L_z/2)=0$ is then \cite{vf2}:
\begin{gather}
g(z)=-\frac{\phi_0}{2\lambda_{ab}^2\sinh(L_z/\lambda_{ab})}\big\{\cosh [(z+u)\lambda_{ab}^{-1}]
\nonumber \\
-\cosh [(L_z-|z-u|)\lambda_{ab}^{-1}] \big\}.
\label{a5}
\end{gather}
Integration of this $g(z)$ in Eq. (\ref{a3}) yields Eq. (\ref{ph})

\end{document}